\PassOptionsToPackage{
    linktocpage,
    pagebackref
}{hyperref}

\documentclass[11pt]{article}

\usepackage{jheppub}

\usepackage[utf8]{inputenc}

\usepackage{amsfonts}

\usepackage{amsmath,amssymb}

\usepackage{mathtools}

\usepackage{multirow}

\usepackage{graphicx} 
\usepackage[space]{grffile}

\usepackage{simplewick}

\usepackage{array}

\usepackage{color}

\usepackage{placeins}
\usepackage{subcaption}

\usepackage{longtable}

\usepackage{dsdshorthand}
\usepackage{mymacro-andrea}

\usepackage{tikz}
\usetikzlibrary{trees}
\usetikzlibrary{decorations.pathmorphing}
\usetikzlibrary{decorations.markings}
\usetikzlibrary{shapes.misc}

\renewcommand*{\backref}[1]{}
\renewcommand*{\backrefalt}[4]{%
  \ifcase #1 %
  \or
    {\scriptsize\,[p.~#2]}%
  \else
    {\scriptsize\,[pp.~#2]}%
  \fi
}

\tikzset{
	threept/.style={
		circle,
		draw,
		inner sep=2pt,
	},
	twopt/.style={
		circle,
		draw,
		fill=black,
		inner sep=1pt,
		minimum size=1pt
	},
	cross/.style={
		cross out,
		draw=black, 
		minimum size=7pt, 
		inner sep=0pt,
		outer sep=0pt
	},
	scalar/.style={
		thick,
		dashed,
		postaction={
			decorate,
			decoration={
				markings,
				mark=at position 0.5 with {\arrow{>}}
			}
		}
	},
	spinning/.style={
		thick,
		postaction={
			decorate,
			decoration={
				markings,
				mark=at position 0.5 with {\arrow{>}}
			}
		}
	},
	spinning no arrow/.style={
		thick,
	},
	finite with arrow/.style={
		decoration={
			snake,
			amplitude=1pt,
			segment length=6pt,
			post length=2pt
		},
		decorate,
		thick,->
	},
	finite/.style={
		decoration={
			snake,
			amplitude=1pt,
			segment length=6pt,
		},
		decorate,
		thick
	}
}

\def\nn{\nonumber}

\usepackage{dsfont}

\definecolor{martenorange}{rgb}{0.91, 0.41, 0.17}

\newcommand{\point}{{\bf p}}

\DeclareMathOperator{\pp}{\bf{p }}

\newcommand{\beq}{\begin{equation}}
\newcommand{\eeq}{\end{equation}}

\newcommand{\structgeneral}{
	\left[
	\begin{matrix}
	q_1 & q_2 & q_3 & q_4 \\
	\bar q_1 & \bar q_2 &\bar q_3 &\bar q_4	
	\end{matrix}
	\right]
}
\newcommand{\struct}[8]{
	\left[
	\begin{matrix}
	#1 & #2 & #3 & #4 \\
	#5 & #6 & #7 & #8	
	\end{matrix}
	\right]
}

\newcommand{\OO}{\mathcal{O}}

\newcommand{\II}{\mathbb{I}}
\newcommand{\JJ}{\mathbb{J}}
\newcommand{\KK}{\mathbb{K}}

\graphicspath{
	{./graphics_Andrea/}
	{./plots}
}

\title{\boldmath Bounds on Abelian Currents in 4d CFTs}

\author[a]{Denis Karateev,}
\author[b]{Petr Kravchuk,}
\author[c]{Andrea Manenti,}
\author[b]{\\Marten Reehorst,}
\author[d]{and Alessandro Vichi}

\affiliation[a]{D\'epartment de Physique Th\'eorique, Universit\'e de Gen\`eve,\\
	24 quai Ernest-Ansermet, 1211 Gen\`eve, Switzerland}
\affiliation[b]{Department of Mathematics, King's College London, Strand, London WC2R 2LS, United Kingdom}
\affiliation[c]{Department of Physics and Astronomy, Uppsala University, Box 516, SE-751 20 Uppsala, Sweden}
\affiliation[d]{Department of Physics, University of Pisa and INFN, Largo Pontecorvo 3, I-56127 Pisa, Italy}

\abstract{
	We study four-dimensional conformal field theories (CFTs) with an abelian $U(1)$ global symmetry using the conformal bootstrap approach. 
	We obtain numerical bounds on the scaling dimensions of low-lying operators, the stress-tensor central charge, and a particular combination of the 't~Hooft anomaly and the current central charge. 
	Our analysis provides the first non-perturbative constraints on four-dimensional CFTs with conserved abelian currents and establishes a framework that can be extended to theories with non-abelian global symmetries.
}

\begin{document}
\maketitle

\section{Introduction}
\label{sec:introduction}

Over the past two decades there has been remarkable progress in the theoretical and numerical understanding of conformal field theories (CFTs) in $d>2$ space-time dimensions, following the seminal work \cite{Rattazzi:2008pe}. For a comprehensive overview of these developments, see the review \cite{Poland:2018epd}.
This progress is particularly striking in three dimensions, where, under a few mild assumptions, several important theories have been solved numerically with extraordinary precision. 
Prominent examples include the numerical solution of the Ising model \cite{El-Showk:2012cjh,El-Showk:2014dwa} and its further precision studies \cite{Kos:2016ysd,Chang:2024whx,Poland:2025ide,Reehorst:2021hmp},
the study of the $O(N)$ models \cite{Kos:2016ysd,Reehorst:2020phk} and their generalisations \cite{Manenti:2021elk,Reehorst:2024vyq},
the precision study of the $O(2)$ model \cite{Chester:2019ifh,Liu:2020tpf}, the precision study of the $O(3)$ model describing Heisenberg magnets \cite{Chester:2020iyt}, and the study of the Gross–Neveu model \cite{Iliesiu:2015qra,Iliesiu:2017nrv,Erramilli:2022kgp,Mitchell:2024hix}. Many interesting results were also obtained in the case of 3d QED and related models in \cite{Chester:2016wrc,Chester:2017vdh,Li:2018lyb,Li:2021emd,He:2021sto,He:2021xvg,Chester:2023njo,Chester:2025uxb}. An account of more recent developments and applications can be found in~\cite{Rychkov:2023wsd}.

While the conformal bootstrap has achieved remarkable results in three dimensions, its application to conformal field theories in four dimensions presents additional challenges and remains an active area of research.
In four dimensions, a variety of bounds on scalar operators, both with and without global symmetries, have been obtained in \cite{Rychkov:2009ij,Rattazzi:2010gj,Rattazzi:2010yc,Poland:2010wg,Vichi:2011ux,Poland:2011ey,Kos:2013tga,Caracciolo:2014cxa,Kos:2015mba}. 
More recently, bounds involving spinning operators, including fermionic operators, have been constructed in \cite{Karateev:2019pvw}. 
Despite this progress, identifying and isolating explicit interacting four-dimensional CFTs remains a central open challenge.

Examples of non-supersymmetric four-dimensional CFTs arise as infrared (IR) fixed points of gauge theories, commonly referred to as Caswell-Banks-Zaks (CBZ) fixed points.
In certain cases, for example in QCD in the Veneziano limit,\footnote{Consider QCD with $N_f$ Dirac fermions in the fundamental representation of the $SU(N_c)$ gauge group, where $N_c$ denotes the number of colours. The Veneziano limit is defined by taking $N_f \to \infty$ and $N_c \to \infty$ while keeping the ratio $N_f/N_c$ finite. One can then introduce the parameter $\varepsilon \equiv \tfrac{11}{2} - N_f/N_c$, which can be made arbitrarily small, providing perturbative control over the theory.} 
these fixed points can be analysed perturbatively, see for example
\cite{Beneke:1997qd,Bauer:1997gs,Baikov:2014qja,Herzog:2017ohr,DiPietro:2020jne} and references therein.
More generally, however, CBZ fixed points are strongly coupled, and their detailed properties are difficult to access. 
At present, the main non-perturbative approach is lattice gauge theory, which nevertheless faces significant challenges in the conformal regime. For an overview of lattice field theory results for the QCD conformal window see for example \cite{DeGrand:2015zxa}. Recent progress includes, among others, \cite{Hasenfratz:2018wpq,Fodor:2018uih,Hasenfratz:2019dpr,Hasenfratz:2024fad}. An alternative controlled approach to QCD-like theories employs AdS backgrounds, where bootstrap methods can be applied and the flat-space limit taken at the end. Recent examples include \cite{Ciccone:2024guw,Ciccone:2025dqx,DiPietro:2025ozw}.

Experience with three-dimensional bootstrap studies shows that it is beneficial to impose in a bootstrap setup as much information about the target theory as possible. The most important piece of information in this regard is the global symmetry of the theory and the associated conserved currents. By focusing on the bootstrap of conserved currents, one gains a much better handle on the symmetry structure and dynamical content of the theory. In 3d conserved currents in various setups have been recently studied in \cite{Dymarsky:2017xzb,Manenti:2018xns,Reehorst:2019pzi,He:2023ewx,Bartlett-Tisdall:2024mbx}.

In four dimensions conserved currents also encode the 't~Hooft anomalies, which are renormalisation-group invariant quantities.  For a given gauge theory, these anomalies can be computed exactly in the ultraviolet, and their values remain unchanged along the RG flow.
In this paper we take a first step towards a systematic study of conserved currents in four-dimensional conformal field theories, concentrating on theories with (at least) an abelian $U(1)$ global symmetry. 
The extension of our analysis to non-abelian global symmetries will be pursued in future work.

\paragraph{Structure of the paper}
We outline the main features of our setup in section \ref{sec:setup}, present our numerical results in section \ref{sec:results}, and discuss their interpretation, limitations, and possible extensions in section \ref{sec:discussion}.

All relevant technical details are collected in the appendices. 
We specify our $d=4$ notation in appendix \ref{app:notation}. 
Appendix \ref{app:three-point_functions} contains the details of the three-point functions, while appendix \ref{app:conformal_blocks} describes the construction of conformal blocks. 
The crossing equations for the current four-point function are discussed in appendix \ref{app:crossing_equations}. 
Finally, we provide details of our numerical setup in appendix \ref{app:numerical_setup}.

\paragraph{Ancillary files}
We also provide several ancillary files accompanying this paper. 
These include our conventions for three-point tensor structures, the associated conformal blocks, and the functionals used to extract independent crossing equations.
Each of these files is described in appendices \ref{app:three-point_functions}--\ref{app:crossing_equations}.

In addition, we provide a Mathematica notebook \verb|paper_plots.nb| containing the numerical data used in the figures of this paper, together with the Mathematica functions used to generate the plots.

The ancillary data for this work are available on Zenodo at
\url{https://doi.org/10.5281/zenodo.18037786}.

\section{Setup}
\label{sec:setup}

The operators in a four-dimensional CFT with a global $U(1)$ symmetry are labelled by their Lorentz representation $(\ell,\bar\ell)$, scaling dimension $\Delta$, and $U(1)$ charge $q$.
In this paper, we study constraints arising from crossing symmetry of the four-point function of conserved currents $\langle JJJJ\rangle$, where $J$ denotes the abelian $U(1)$ current.
Since we focus exclusively on this correlator, we only encounter neutral operators with $q=0$.
Throughout, we adopt the convention that $(0,0)$ denotes a scalar operator, $(1,0)$ a left-handed fermion, and $(1,1)$ a vector.
We also assume parity invariance. Parity relates operators in the Lorentz representation $(\ell,\bar\ell)$ to those in $(\bar\ell,\ell)$.

\paragraph{Two-point functions}
Most two-point correlation functions of local operators are completely fixed by conformal invariance once normalization conventions are specified. 
The only exceptions are the two-point functions of abelian conserved currents $J$ and of the stress tensor $T$, whose normalizations are fixed by Ward identities associated with the global $U(1)$ symmetry and conformal symmetry, respectively.
For example, the conserved current $J^\mu(x)$ is canonically normalized so that the charge
$
Q=\int d^3x\, J^0(x)
$
is properly quantized. This normalization is reflected in Ward identities for correlation functions involving operators with nonzero $U(1)$ charge $q\neq 0$.

The normalization of $J$ and $T$ is encoded in the real dimensionless parameters
\begin{equation}
	C_J \qquad \text{and} \qquad C_T,
\end{equation}
called the current and stress-tensor central charges respectively.
These parameters, together with conformal invariance completely determine the two-point functions $\langle J J \rangle$ and $\langle T T \rangle$.

Since in this work we are able to only study the neutral sector of the theory, the canonical normalization of the current cannot be imposed directly, and all physical observables must be expressed in units of $C_J$.
From the definition of the current central charge, the rescaled two-point function $C_J^{-1}\langle J J \rangle$ is independent of $C_J$, which motivates working with the rescaled current $C_J^{-1/2} J$ throughout.
Consequently, rather than constraining the three- or four-point function $\langle J J J \rangle$ and $\langle J J J J \rangle$ themselves, our bootstrap analysis yields bounds on the rescaled current correlators
\begin{equation}
	\label{eq:3pt_rescaled_example}
	C_J^{-3/2}\langle J J J \rangle,\qquad
	C_J^{-2}\langle J J J J \rangle.
\end{equation}

\paragraph{Three-point functions}
Three-point functions of local operators are fixed by conformal invariance up to numerical coefficients, known as the OPE coefficients. To analyse the four-point correlator of conserved currents $\langle JJJJ\rangle$, one must first understand the structure of three-point functions involving two currents and a general operator $\cO$, schematically written as $\langle JJ\cO\rangle$. These correlators are non-vanishing only when $\cO$ transforms in one of the following Lorentz representations:
\begin{equation}
	(\ell_\cO, \bar\ell_\cO): \qquad
	(\ell,\ell),\quad
	(\ell,\ell+2),\quad
	(\ell+2,\ell),\quad
	(\ell,\ell+4),\quad
	(\ell+4,\ell),
\end{equation}
where $\ell = 0,1,2,\ldots$ 
For each spin representation $(\ell, \bar\ell)$ there exists an infinite tower of primary operators. We denote by $\Delta_{(\ell,\bar\ell)}$ the lowest scaling dimension in this tower, by $\Delta'_{(\ell,\bar\ell)}$ the next-to-lowest one, and so on. For instance, the tower of scalar operators is characterised by the sequence of scaling dimensions
\begin{equation}
	\Delta_{(0,0)} \;<\; \Delta'_{(0,0)}  \;<\; \Delta''_{(0,0)} \;<\;\ldots
\end{equation}

Some three-point functions play a particularly important role. The three-point function $\langle JJJ \rangle$ has a single parity-odd tensor structure, and we denote the OPE coefficient multiplying this structure by $\lambda^-_{JJJ}$. 
This OPE coefficient is related to the $U(1)^3$ ’t~Hooft anomaly \cite{tHooft:1979rat}. Consider a generic free theory consisting of fundamental scalars, vector fields, and Weyl fermions. Only the fermions contribute to the ’t~Hooft anomaly. Denoting the $U(1)$ charges of the free Weyl fermions\footnote{Recall that we can take all fundamental Weyl fermions to be left-handed by convention.} by $q_i$, the OPE coefficient $\lambda^-_{JJJ}$ in such a generic free theory is given by
\begin{align}\label{eq:general_JJJ}
	\lambda^-_{JJJ}=\frac{i}{2\sqrt{2}\pi^6}\sum_i q_i^3.
\end{align}
The coefficient $\lambda^-_{JJJ}$ takes the above value in all IR fixed points that can be reached by $U(1)$-preserving RG flows from this free theory in the UV, provided the current $J$ is identified appropriately.

There are two parity-even tensor structures describing the $\langle JJT \rangle$ correlation function. The two OPE coefficients multiplying these structures are denote by $\lambda^+_{JJT,\,1}$ and $\lambda^+_{JJT,\,2}$.
They are constrained by the following Ward identity for $T$,
\begin{equation}
	\label{eq:ward_identities_JJT}
	\lambda^+_{JJT,\,1} = \frac{4C_J}{\pi^2},
\end{equation}
where $C_J$ is the current central charge introduced earlier. 
This and other Ward identities for spinning operators can be efficiently derived using the technology described in appendix~B of \cite{Karateev:2019pvw}. 
Additional examples of four-dimensional spinning Ward identities, including correlators involving fermions and the stress tensor, can be found in \cite{Elkhidir:2017iov,Karateev:2019pvw}.
It is convenient to parametrize the OPE coefficients $\lambda^+_{JJT,\,1}$ and $\lambda^+_{JJT,\,2}$ in terms of the parameter $\gamma$ as follows
\begin{equation}
	\label{eq:representation_JJT_coefficients}
	\lambda^+_{JJT,\,1} =  \frac{4C_J}{\pi^2},\qquad
	\lambda^+_{JJT,\,2} = - \frac{8C_J}{9\pi^2}\left(1+4 \gamma\right).
\end{equation}
This parametrization automatically satisfies \eqref{eq:ward_identities_JJT}. Furthermore, the Hofman–Maldacena bounds \cite{Hofman:2008ar} imply
\begin{equation}
	\label{eq:HM_bound}
	-\frac{1}{16} \leq \gamma \leq \frac{1}{32},
\end{equation}
with the lower bound saturated by the free massless complex scalar and the upper bound by the free Weyl fermion.

\paragraph{Unitarity bounds}

In unitary CFTs in four dimensions, the scaling dimension of any  primary operator must  obey the unitarity bounds \cite{Mack:1975je}
\begin{align}
& \Delta_{(\ell+p, \ell)} \geq \ell + \frac{p}2 +2,\qquad  \ell\neq 0\, , \nonumber\\
& \Delta_{(p, 0)} \geq  \frac{p}2 +1 \, .
\end{align}
and the same relations hold for $\Delta_{(\ell, \ell+p)}$ and $\Delta_{(0, p)}$. In \cite{Cordova:2017dhq,Manenti:2019kbl}, using the averaged null energy condition (ANEC), the authors derived a stronger bound on the scaling dimension of chiral operators, of the form $ \Delta_{(p, 0)}\geq p$ (for $p\leq 20$). The latter bound coincides with the unitarity bounds for $p=2$ and it is stronger for larger values of $p$. A bound for  $\Delta_{(p+1, 1)}$ also exists, but it is not relevant for this work. It is plausible that similar bounds exist for generic $\Delta_{(\ell+p, \ell)} $ when $p$ is sufficiently large. We will not make use of ANEC bounds in the present work. 

\paragraph{Free complex scalar}

The simplest CFT in four dimensions consists of a single free complex scalar $\phi(x)$ with all local operators constructed as polynomials in $\phi$ and $\phi^\dagger$, and their derivatives. The neutral scalar of lowest scaling dimension is $ \phi\phi^\dagger$. 
Constructing chiral operators is more subtle, since the simplest candidates are either descendants or vanish identically due to antisymmetrization of Lorentz indices. The lowest-dimension scalar operator in the $(4,0)$ representation requires four fields and four derivatives. Hence
\begin{equation}
	\label{eq:scaling_dimensions_CS}
	\Delta_{(0,0)} = 2,\qquad
	\Delta_{(4,0)} =8.
\end{equation}

The central charges of the conserved currents and the stress-tensor, assuming $\phi$ has $U(1)$ charge $1$, are
 \begin{equation}
 	\label{eq:weyl_central_charges}
 	C_J = \frac{1}{8\pi^4},\qquad
 	C_T = \frac{2}{3\pi^4}.
 \end{equation}
By computing the three-point functions we can extract the following OPE coefficients
\begin{equation}
	\label{eq:weyl_fermion_results}
	\lambda^-_{JJJ} = 0,\qquad
	\lambda^+_{JJT,\,1} = \frac{1}{2\pi^6},\qquad
	\lambda^+_{JJT,\,2} = - \frac{1}{12\pi^6}.
\end{equation}
These agree with~\eqref{eq:general_JJJ} and \eqref{eq:representation_JJT_coefficients} if we set $\gamma=-1/16$.

\paragraph{Free Weyl fermion}
In the theory of a free left-handed Weyl fermion $\psi_\alpha(x)$, all local operators are constructed as polynomials in $\psi$ and its derivatives. Thus, we can infer their scaling dimensions. In particular, the lightest scalar operator and the lightest $(4,0)$ operator have the following scaling dimensions
\begin{equation}
	\label{eq:scaling_dimensions_WF}
	\Delta_{(0,0)} = 6,\qquad
	\Delta_{(4,0)} = 8.
\end{equation}

The central charges of the conserved currents and the stress-tensor are given by, assuming $\psi$ has $U(1)$ charge $1$,
 \begin{equation}
 	\label{eq:weyl_central_charges}
 	C_J = \frac{1}{2\pi^4},\qquad
 	C_T = \frac{1}{\pi^4}.
 \end{equation}
By computing the three-point functions we can extract the following OPE coefficients
\begin{equation}
	\label{eq:weyl_fermion_results}
	\lambda^-_{JJJ} = \frac{i}{2\sqrt{2}\pi^6},\qquad
	\lambda^+_{JJT,\,1} = \frac{2}{\pi^6},\qquad
	\lambda^+_{JJT,\,2} = - \frac{1}{2\pi^6}.
\end{equation}
These agree with~\eqref{eq:general_JJJ} and \eqref{eq:representation_JJT_coefficients} if we set $\gamma=1/32$ and take into account \eqref{eq:weyl_central_charges}.

\paragraph{Crossing equations}

Our bounds are based on the crossing symmetry of the four-point function $\<JJJJ\>$. This four-point function contains $70$ independent tensor structures. 
In this work we impose parity invariance, which removes $27$ parity-odd structures, leaving $43$ parity-even ones. 
Furthermore, since all abelian currents are identical, permutation symmetry further reduces this number to $19$ tensor structures. 
A priori, each of these $19$ structures leads to a crossing equation. 
However, due to current conservation this is not the case: in practice, only $7$ of them yield independent crossing equations in our case. All $7$ equations are two-variable equations, we do not find any line or point equations.
Using an independent set of equations is crucial for ensuring numerical stability. 
A detailed discussion of the crossing equations for abelian currents is presented in appendix \ref{app:crossing_equations}.

\paragraph{Numerical bounds}

\begin{figure}[t] 
	\centering
	\includegraphics[width=0.75\linewidth]{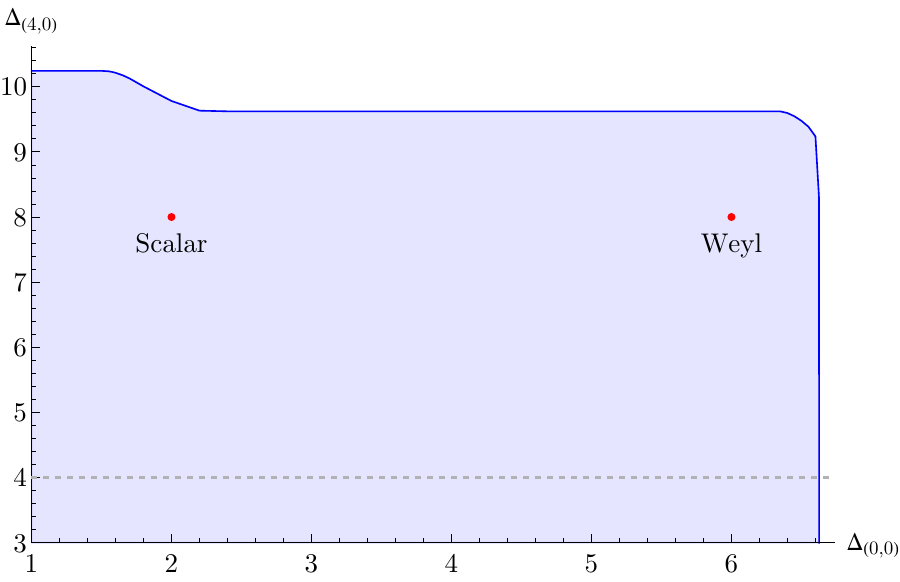}
	\caption{Bound on the scaling dimension $\Delta_{(4,0)}$ as a function of $\Delta_{(0,0)}$, at $n_\text{max}=16$. The allowed region lies below the blue curve. The red dots indicate the locations of free complex scalar and free Weyl fermion theories. The dashed line indicates the analytic lower bound $\Delta_{(4,0)}\geq 4$ found in \cite{Cordova:2017dhq} using the ANEC.}
	\label{fig:bound_scaling_dimensions}
\end{figure}

Using the numerical conformal bootstrap, one can place bounds on scaling dimensions and on the coefficients appearing in the conformal block decomposition of the four-point function $\langle JJJJ\rangle$. As an illustrative example, in this work we focus on the following two scaling dimensions:
\begin{equation}
	\Delta_{(0,0)} \qquad \text{and} \qquad \Delta_{(4,0)}.
\end{equation}
Our numerical bound on these scaling dimensions can be found in figure~\ref{fig:bound_scaling_dimensions}.

We denote by $-P_J$ the coefficient appearing in front of the conformal block encoding the contribution of the conserved current $J$ exchange.\footnote{Note that in our conventions $\l_{JJJ}^-$ is pure imaginary, which is why the coefficient of $J$ block is negative.} This coefficient is given by the following ratio of the 't Hooft anomaly and the current central charge
\begin{equation}
	\label{eq:P_J}
	P_J \equiv \frac{|\lambda^{-}_{JJJ}|^2}{C_J^3}.
\end{equation}
For a free left-handed Weyl fermion one finds $P_J = 1$, as follows from 
\eqref{eq:weyl_central_charges} and \eqref{eq:weyl_fermion_results}. 
Note that in a general free theory with left-handed Weyl fermions of charges $q_i$ 
and any additional bosonic matter, we have
\begin{align}
	C_J \geq \frac{1}{2\pi^4}\sum_i q_i^2 ,
\end{align}
while $\lambda^-_{JJJ}$ is given by \eqref{eq:general_JJJ}. 
This bound is saturated by the free theory consisting only of Weyl fermions as can be seen from \eqref{eq:weyl_central_charges}. 
The contribution from bosonic operators in free theories was computed in \cite{Osborn:1993cr} and is always non-negative; hence we obtain the above bound.
This implies, in free theories,
\begin{align}
	\label{eq:bound_free_theories}
	P_J \leq \frac{\p{\sum_i q_i^3}^2}{\p{\sum_i q_i^2}^3}
	\leq \p{\frac{\p{\sum_i |q|_i^3}^{1/3}}{\p{\sum_i q_i^2}^{1/2}}}^6\leq 1.
\end{align}
Therefore, the theory of a single Weyl fermion maximizes the value of $P_J$ among free theories. Our numerical bound on $P_J$ can be found in figure~\ref{fig:convergence}.

Related bounds on the ratio of the $U(1)$ ’t~Hooft anomaly to the current two-point normalization — corresponding to $\sqrt{P_J}$ in our notation up to convention-dependent factors — have been obtained in the context of four-dimensional $\mathcal N=1$ superconformal field theories in \cite{Lin:2019vgi}.

We denote by $P_T$ the coefficient appearing in front of the conformal block encoding the contribution of the stress-tensor $T$ exchange. It is precisely defined by
\begin{equation}
	\label{eq:P_T}
	P_T^{ab} \equiv \frac{	\lambda^+_{JJT,\,a} \lambda^+_{JJT,\,b}  }{C_J^2 C_T}.
\end{equation}
The $2\times2$ matrix $P_T$ is symmetric and positive semi-definite, $P_T \succeq 0$.
Using the parametrisation~\eqref{eq:representation_JJT_coefficients}, it can be written explicitly as
\begin{equation}\label{eq:PT}
	P_T = \frac{16}{81\pi^4 C_T}\;
	\begin{pmatrix}
		81 && -18\,(1+4\gamma)\\
		-18\,(1+4\gamma) && 4\,(1+4\gamma)^2
	\end{pmatrix}.
\end{equation}
We can put bounds on $P_T$ as a function of $\gamma$. These can be translated as bound on $C_T$ as a function of $\gamma$. Our numerical bounds of $C_T$ with various assumptions can be found in figures  \ref{fig:bound_CT_part_1} and \ref{fig:bound_CT_part_2}.

Some of the results presented here were anticipated in the PhD thesis \cite{Manenti:2020nyv}, which also contains additional material and exploratory plots (see figures 8.1 and 8.2) obtained using less computing power.

\section{Numerical results}
\label{sec:results}

Let us now present our numerical results. They are obtained by analysing the crossing equations for the four-point correlation function $\langle JJJJ\rangle$.
In order to do this, one requires an efficient method for approximating spinning conformal blocks in four dimensions. This problem was solved in~\cite{Karateev:2019pvw}, and we follow their strategy in our implementation.
The crossing equations are then studied numerically using the semidefinite programming solver SDPB~\cite{Simmons-Duffin:2015qma,Landry:2019qug}, including the polynomial interpolation optimizations of~\cite{Chang:2025mwt}.

The functionals used in the numerical bootstrap method are constructed using 
$\partial_z^m \partial_{\bar z}^n$ derivatives satisfying $m+n \leq \Lambda$, 
where $\Lambda$ is usually taken to be odd. Instead of $\Lambda$, one often uses 
another integer parameter $n_\text{max}$, related to $\Lambda$ via $\Lambda = 2n_\text{max}-1$. 
All our bounds are rigorous; 
however, they improve as $n_\text{max}$ is increased. 
Other parameter choices used in our numerics, including the SDPB-parameters and the parameters controlling the conformal block approximations, are discussed in appendix \ref{app:numerical_setup}.

\begin{figure}[t] 
	\centering
	\includegraphics[width=0.85\linewidth]{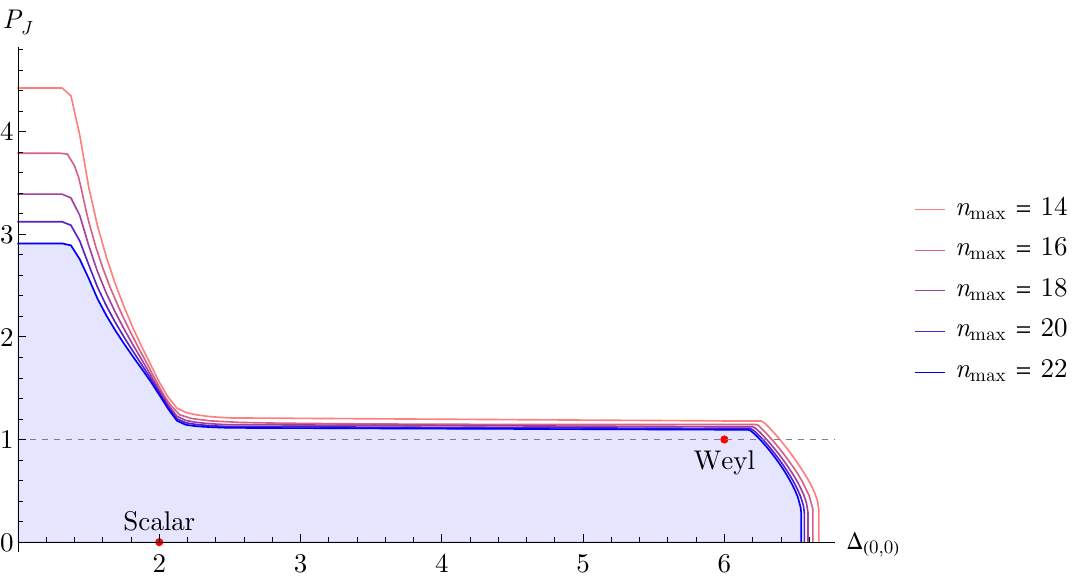}
	\caption{Upper bound on the coefficient $P_J \equiv C_J^{-3}|\lambda^-_{JJJ}|^2$, defined in \eqref{eq:P_J}, as a function of $\Delta_{(0,0)}$. Different curves correspond to different values of $n_\text{max}$. The horizontal dashed line represents the analytic upper bound $P_J\leq 1$ for free theories. Free theories must lie below this line. Above this line the theory cannot be free and must be interacting.
	}
	\label{fig:convergence}
\end{figure}

Our first result is displayed in figure~\ref{fig:bound_scaling_dimensions}. It provides the allowed region of the scaling dimension $\Delta_{(4,0)}$ as a function of the scaling dimension $\Delta_{(0,0)}$. The allowed region lies below the blue curve. This bound exhibits three particularly noticeable kinks, located at
\begin{equation}
	\label{eq:kinks_1}
	\left(\Delta_{(0,0)}, \Delta_{(4,0)}\right):\qquad
	(1.3,\,10.2),\quad
	(2.2,\, 9.6),\quad
	(6.3,\, 9.6).
\end{equation}
The free fermion theory $(6,8)$ lies relatively closely to the right-most kink. The unitarity bounds on the scaling dimensions appearing in this plot read as follows
\begin{equation}
	\Delta_{(0,0)}\geq 1,\qquad
	\Delta_{(4,0)}\geq 3.
\end{equation}
In \cite{Cordova:2017dhq,Manenti:2019kbl}, using the averaged null energy condition (ANEC), the authors derived a stronger ``unitarity'' bound on the scaling dimension of the $(4,0)$ operator. Their analytic bound reads $\Delta_{(4,0)} \geq 4$ and is indicated by the dashed line in figure~\ref{fig:bound_scaling_dimensions}. In our numerical setup, we assume only the standard unitarity bounds ($\Delta_{(4,0)}$ is the only dimension in our setup that could be affected by this).

Note that we do not include the free vector theory (free Maxwell theory) in our plots. 
A free real vector field does not possess a global $U(1)$ symmetry. 
While a theory of a complex free vector field does admit a global $U(1)$ symmetry, it nevertheless does not contain a corresponding gauge-invariant conserved $U(1)$ current operator.\footnote{This does not violate Noether’s theorem: although a Noether current exists, it is not gauge invariant. This occurs because the global $U(1)$ symmetry does not commute with gauge transformations.}

\begin{figure}[t] 
	\centering
	\includegraphics[width=0.9\linewidth]{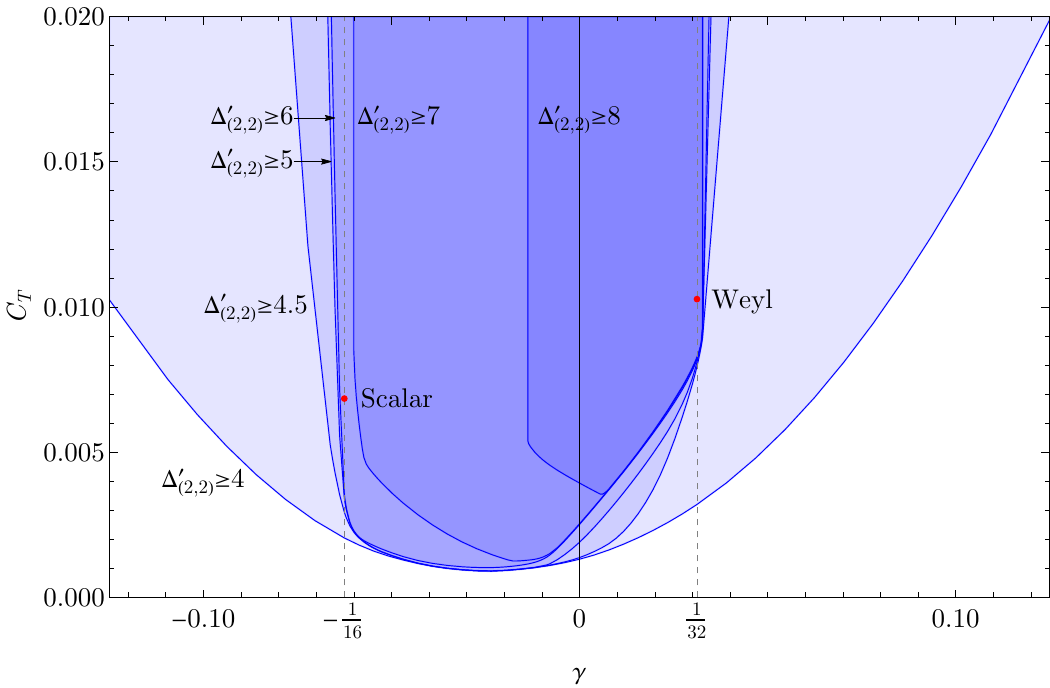}
	\caption{Lower bound on the stress-tensor central charge $C_T$ as a function of $\gamma$, at $n_\text{max}=16$. The different allowed regions correspond to different assumed values of the gap to the second lightest spin-2 operator after the stress-tensor $\Delta'_{(2,2)}$, ranging over $\Delta'_{(2,2)} = \{4,\, 4.5\,, 5,\, 6,\, 7,\, 8\}$. The two dashed vertical lines indicate the Hofman–Maldacena bounds on $\gamma$, given in~\eqref{eq:HM_bound}. The red dots show the free complex scalar and the free Weyl fermion theories. The gaps are $\De'_{(2,2)}=6$ in free complex scalar and $\De'_{(2,2)}=8$ in free Weyl fermion.}
	\label{fig:bound_CT_part_1}
\end{figure}

Our second result is displayed in figure~\ref{fig:convergence}. It shows the allowed region for the coefficient $P_J$ defined in \eqref{eq:P_J} as a function  of the scaling dimension $\Delta_{(0,0)}$. It also shows the dependence of the bound on $n_\text{max}$. Recall that the coefficient $P_J$ is sensitive to the 't Hooft anomaly. Our bound again exhibits three particularly noticeable kinks. They appear at the following positions
\begin{equation}
	\label{eq:kinks_2}
	\left(\Delta_{(0,0)}, P_J\right):\qquad
	(1.5,\,2.9),\quad
	(2.1,\, 1.2),\quad
	(6.2,\, 1.1).
\end{equation}
The $\Delta_{(0,0)}$ locations of these kinks are approximately the same as those in \eqref{eq:kinks_1}. The free fermion theory $(6,1)$ again lies relatively close to the right-most kink. According to equation \eqref{eq:bound_free_theories}, all free theories obey $P_J \leq 1$. This analytic bound is indicated by the dotted horizontal line, and all free theories lie below it. As a consequence, all theories above this line must be interacting.

As far as the dependence on $n_\text{max}$ goes, the bound has mostly converged at $\De_{(0,0)}\gtrsim2.2$, but is far from being converged below $\De_{(0,0)}=2$. We have not made a systematic attempt at extrapolating this bound, but a superficial analysis suggests that anything in between roughly $P_J\leq 1$ and $P_J\leq 2$ is a reasonable possibility for the $n_\text{max}=\oo$ bound at the gap $\De_{(0,0)}=1$.

\begin{figure}[t] 
	\centering
	\includegraphics[width=0.9\linewidth]{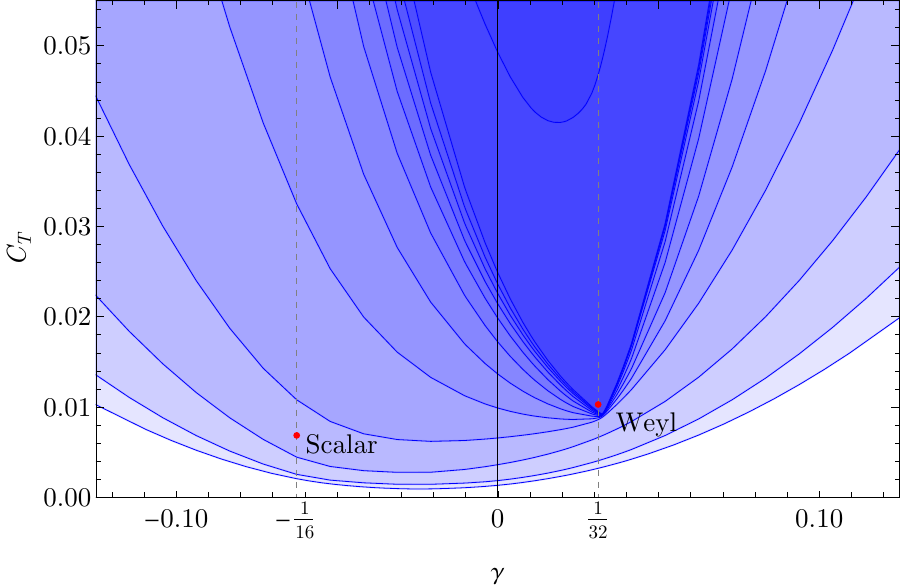}
	\caption{Lower bound on the stress-tensor central charge $C_T$ as a function of $\gamma$, at $n_\text{max}=16$. The different allowed regions correspond to different gaps $\Delta_{(0,0)}$ to the lightest scalar, ranging over
	the set 
	$
		\{1.0,\, 1.5\,, 1.96,\, 2.42,\, 2.88,\, 3.33,\, 3.79,\, 4.25,\, 4.71,\, 5.17,\, 5.63,\, 6.08,\, 6.54\}.
	$
	The	\mbox{smallest} values of $\Delta_{(0,0)}$ give the weakest bounds. The two dashed vertical lines indicate the Hofman-Maldacena bounds on $\gamma$ given in~\eqref{eq:HM_bound}. 
	The red dots denote the free complex scalar and the free Weyl fermion theories. 
	}
	\label{fig:bound_CT_part_2}
\end{figure}

Finally, we present our lower bounds on the stress-tensor central charge $C_T$ in figures \ref{fig:bound_CT_part_1} and \ref{fig:bound_CT_part_2}.
In both cases the allowed region lies above the coloured curves, which correspond to different additional assumptions.
In figure~\ref{fig:bound_CT_part_1} we fix the gap $\Delta'_{(2,2)}$ to the next spin-two operator after the stress-tensor, $\Delta_{(2,2)} = 4$, while in figure \ref{fig:bound_CT_part_2} we vary the gap $\Delta_{(0,0)}$ to the lightest scalar. 

Note that if one does not impose a gap $\Delta'_{(2,2)}>4$ above the spin-2 unitarity bound, additional conserved spin-2 primaries at $\Delta=4$ can contribute alongside $T$ (e.g.\ as found in product CFTs). In this case we cannot guarantee that the operator we call $T$ and whose contribution~\eqref{eq:PT} we isolate in the numerics is the full stress-tensor operator, or even a positive linear combination of decoupled stress-tensor operators. In this situation the Hofman--Maldacena bounds on $\gamma$ need not apply.\footnote{Explicilty, an arbitrary positive-semidefinite matrix $M$, coming from the limit of $\De>4$ continuum, is effectively added to the contribution $P_T$ defined in~\eqref{eq:PT}. If our crossing equations can detect Hofman-Maldacena bounds at all, they can only impose them on the sum $M+P_T$. This cannot be stronger than requiring that $M+P_T$ be a sum of (possibly multiple, since we cannot exclude decoupled theories) stress-tensor contributions of the form~\eqref{eq:PT}, each satisfying Hofman-Maldacena bounds. But it is not hard to show that $P_T$ in $M+P_T$ is not constrained by this requirement in any way.} Indeed, in figure \ref{fig:bound_CT_part_1} we see that numerically these bounds are only recovered at a sufficiently large gap $\Delta'_{(2,2)}>4$. This behavior is similar to what is found in $\<JJJJ\>$ bootstrap in three dimensions~\cite{Dymarsky:2017xzb}. Note that this is different from $\<TTTT\>$ bootstrap in three dimensions~\cite{Dymarsky:2017yzx,Chang:2024whx}, where the $T$ contribution is effectively isolated by tensor structure counting ($\<TTT\>$ has more tensor structures than $\<TT\cO\>$ for generic spin-2 $\cO$) and Hofman--Maldacena bounds are reproduced without additional gaps.

It is noteworthy that for $\Delta_{(0,0)} \approx 6$ the lower bound in figure~\ref{fig:bound_CT_part_2} accumulates near the free Weyl fermion point (shown as a red dot), providing a strong consistency check of our results.

\FloatBarrier

\section{Discussion}
\label{sec:discussion}

One of the most interesting aspects of studying conserved currents is the fact that we gain access to the 't~Hooft anomaly, an RG-invariant quantity. 
In the abelian case, however, we do not probe the anomaly directly. 
Instead, we constrain the combination $P_J \equiv C_J^{-3}|\lambda^-_{JJJ}|^2$, which depends not only on the 't~Hooft anomaly but also on the current central charge $C_J$, the latter being non--RG-invariant. Consequently, $P_J$ itself is not RG-invariant. 
The resulting bound on $P_J$ was presented in figure~\ref{fig:convergence}.  In future work we plan to study the much more involved case of non-abelian currents, where it becomes possible to disentangle $C_J$ from the 't~Hooft anomaly and, thus, to probe the 't~Hooft anomaly directly. Another class of setups that can provide direct access to $\lambda^-_{JJJ}$ even in the abelian case involves mixed systems containing both conserved currents and charged operators.

One of the main challenges of this work is controlling the complexity associated with the kinematics of spinning operators in four dimensions. 
The problem of constructing two-, three- and four-point correlation functions in $d=4$ was fully solved in \cite{Dolan:2000ut,Dolan:2003hv,Dolan:2011dv}, \cite{Costa:2011mg,Costa:2011dw} and  \cite{Simmons-Duffin:2012juh,Elkhidir:2014woa,CastedoEcheverri:2015mkz,Costa:2016hju,CastedoEcheverri:2016dfa,Cuomo:2017wme,Karateev:2017jgd}.
Moreover, this technology has been implemented in the \texttt{CFTs4D} Mathematica package \cite{Cuomo:2017wme}, which enables one to perform the symbolic manipulations that would otherwise be impossible by hand. 
Despite this progress, the bookkeeping involved in four-dimensional CFTs remains nontrivial. 
This issue is addressed in the forthcoming work \cite{Karateev:2026xxx}, where a systematic solution to the bookkeeping problem is presented. 
Details on the kinematics of the abelian current setup and the associated bookkeeping can be found in appendices on three-point functions \ref{app:three-point_functions} and conformal blocks \ref{app:conformal_blocks}.

The $U(1)$-symmetric setup studied in this work should be viewed as a technical starting point towards the study of more realistic global symmetry structures relevant for four-dimensional gauge theories. For example, standard QCD, a gauge theory with $N_f$ Dirac fermions in the fundamental representation of the $SU(N_c)$ gauge group, possesses an $\SU(N_f)\times\SU(N_f)\times U(1)$ global symmetry; while adjoint QCD, a  gauge theory with $N_f$ Weyl fermions in the adjoint representation of the $SU(N_c)$ gauge group, has an $\SU(N_f)$ global symmetry. Extending the present analysis to CFTs with $\SU(N_f)$ global symmetry is currently in progress.
At the same time, the abelian symmetry studied here can be regarded as an abelian subgroup of a larger non-abelian flavour symmetry appearing in QCD-like theories. This feature is both a strength and a limitation of our approach: the resulting bounds are very general and apply to a wide class of theories, but precisely because of this generality they are not expected to isolate any specific four-dimensional CFT.

The technology developed in this paper can also be applied to four-dimensional $\mathcal{N}=1$ supersymmetric conformal field theories. Any such theory necessarily contains a conserved $U(1)_R$ symmetry. The corresponding $R$-current is not an isolated operator: it is the superconformal primary of the Ferrara–Zumino multiplet, which also contains the stress tensor and the supersymmetry currents. As a result, imposing supersymmetry tightly correlates the dynamics of conserved currents, the stress tensor, and supercharges, providing additional structure that can be exploited in bootstrap analyses.
$\mathcal{N}=1$ supersymmetry can be incorporated into our setup by replacing ordinary conformal blocks with $\mathcal{N}=1$ superconformal blocks. In the present context, the relevant superconformal blocks can be written as linear combinations of the conformal blocks computed in this paper. The coefficients in these linear combinations are fixed by representation theory and have been computed in  \cite{Manenti:2018xns}. 

On the other hand, global symmetry currents in four-dimensional $\mathcal{N}=1$ theories belong to a scalar supermultiplet, while the vector current is a $Q\bar Q$ descendant. Owing to this simplification, numerical bootstrap studies involving the scalar superprimary have already been attempted, but the resulting bounds are still far from the known results \cite{Li:2017ddj,Lin:2019vgi}. One possible improvement is to perform a mixed-correlator study involving both the scalar superprimary and the vector descendant. Once again, the relations among the various OPE coefficients are fixed by representation theory and can be obtained straightforwardly using the results of \cite{Manenti:2019jds}.

Another natural extension of this work is the study of crossing symmetry of the stress-tensor four-point function $\<TTTT\>$ in four dimensions. In three-dimensions this correlator has been first studied in \cite{Dymarsky:2017yzx} and later in \cite{Chang:2024whx}. Based on the experience from these works, one can expect to obtain interesting bounds on the space of all four-dimensional CFTs and be able to access bounds on the Weyl anomaly $a$, which is monotonic under RG flows \cite{Cardy:1988cwa,Osborn:1989td,Jack:1990eb,Komargodski:2011vj,Luty:2012ww}.

\section*{Acknowledgments}
We are grateful to Fabiana De Cesare, Shai Chester, Rajeev Erramilli, Colum Flynn, Yu Nakayama, David Poland, Slava Rychkov, David Simmons-Duffin, Andy Stergiou and Emilio Trevisani for useful conversations. We also thank Andy Stergiou for the initial collaboration on this project.

This work was presented at the Yukawa Institute for Theoretical Physics at Kyoto University during the workshop Progress of Theoretical Bootstrap. We thank the organizers and participants for stimulating discussions.

PK would like to acknowledge the support from the Institute for Advanced study and the Simons Center for Geometry and Physics, where parts of this work were completed.

The work of DK is funded by the Swiss State Secretariat for Education, Research and Innovation (SERI) under contract number MB25.00001 and by the SNSF Ambizione grant PZ00P2\_193411. 

The work of PK and MR is funded by UK Research and Innovation (UKRI) under the UK
government's Horizon Europe funding Guarantee [grant number EP/X042618/1]. The work of PK is funded
Science and Technology Facilities Council [grant number ST/X000753/1].

AV has received funds from the Swiss National Science Foundation  [grant
no. PP00P2-163670] and from the European Research Council Starting Grant  [grant no. 758903].

The calculations in this work have been performed on CREATE High Performance Cluster~\cite{CREATE} at King's College London and on the EPFL SCITAS cluster.

\appendix

\section{Conventions}
\label{app:notation}
We work in four-dimensional Minkowski space with a mostly-plus metric signature. 

The conserved current associated with the $U(1)$ global symmetry is denoted by $J^\mu(x)$ and satisfies the standard conservation equation
\begin{equation}
	\label{eq:conservation_current}
	\partial_\mu J^\mu(x) = 0.
\end{equation}
It is often convenient to introduce an index-free representation of this operator. 
In the vector formalism of \cite{Costa:2011mg}, it is written as
\begin{equation}
	\label{eq:J_vector}
	J(x,z) \equiv J_\mu(x)\, z^\mu,
\end{equation}
where $z^\mu$ is an auxiliary polarization vector satisfying $z^2 = 0$. 
Analogously, any traceless symmetric tensor operator can be encoded in an index-free form $\cO(x,z)$.

\paragraph{Spinor notation}
We define the conserved current with spinor indices as
\begin{equation}
	J^{\dot\alpha \beta}(x) \equiv -\frac{1}{\sqrt{2}}\,\bar \sigma^{\mu\, \dot\alpha \beta} J_\mu(x).
\end{equation}
In this representation one can again employ an index-free formalism, as developed in \cite{SimmonsDuffin:2012uy,Elkhidir:2014woa}. A systematic summary of this approach is given in \cite{Cuomo:2017wme}, whose conventions we adopt throughout this work.
We introduce
\begin{equation}
	\label{eq:J_spinor}
	J(x,s,\bar s) \equiv J_{\beta}^{\dot\alpha}(x)\,
	s^\beta \bar s_{\dot\alpha},
\end{equation}
where $s_\alpha$ and $\bar s^{\dot\alpha}$ are auxiliary spinor polarizations. 
We follow the conventions of appendix~A in \cite{Karateev:2019pvw}, where the vector and spinor polarizations are related by
\begin{equation}
	z^\mu = -2^{-1/2}\,(\bar s \bar \sigma^\mu s).
\end{equation}
In this convention the following identity holds:
\begin{equation}
	J(x,z) = J(x,s,\bar s).
\end{equation}

Let us now consider a generic operator with scaling dimension $\Delta$ and Lorentz representation $(\ell, \bar\ell)$, denoted by $\OO_\Delta^{(\ell, \bar \ell)}(x,s,\bar s)$. 
To express such operators compactly we again use the spinor index formalism, analogous to \eqref{eq:J_spinor}. 
In this notation one writes
\begin{equation}
	\label{eq:index_free formalism}
	\OO_\Delta^{(\ell, \bar \ell)}(x,s,\bar s) \equiv  
	\left(\OO_\Delta\right){}^{\dot\alpha_1\ldots\dot\alpha_{\bar\ell}}_{\beta_1\ldots\beta_{\ell}}(x)
	\left(s^{\beta_1}\ldots s^{\beta_\ell}\right)
	\left(\bar s_{\dot\alpha_1}\ldots \bar s_{\dot\alpha_{\bar\ell}}\right).
\end{equation}
It is also convenient to introduce the shorthand notation
\begin{equation}
	\OO_\Delta^{(\ell, \bar \ell)}(\point) \equiv  \OO_\Delta^{(\ell, \bar \ell)}(x,s,\bar s),
\end{equation}
where $\point$ collectively denotes the $(x,s,\bar s)$ dependence.

\paragraph{Hermitian conjugation}
Under hermitian conjugation the dotted indices become undotted and vice versa. We require that the $s$ and $\bar s$ polarizations are related by hermitian conjugation, thus
\begin{equation}
	\label{eq:relation_s_sbar}
	\left(s^\beta\right)^\dagger = \bar s^{\dot\beta}
	\qquad\text{and}\qquad
	\left(\bar s_{\dot\alpha}\right)^\dagger = s_{\alpha}.
\end{equation}
The operator in the $(\ell,\bar \ell)$ Lorentz representation under hermitian conjugation behaves as
\begin{equation}
	\left(\OO_\Delta{}^{\dot\alpha_1\ldots\dot\alpha_{\bar\ell}}_{\beta_1\ldots\beta_{\ell}}(x)\right)^\dagger =
	\left(\OO^\dagger_\Delta\right){}^{\alpha_1\ldots\alpha_{\bar\ell}}_{\dot\beta_1\ldots\dot\beta_{\ell}}(x).
\end{equation}
In other words this operator transforms in the $(\bar \ell, \ell)$ representation of the Lorentz group. We can also apply the hermitian conjugation to \eqref{eq:index_free formalism}, the result reads
\begin{equation}
\begin{aligned}
		\left(\OO_\Delta^{(\ell, \bar \ell)}(x,s,\bar s) \right)^\dagger &= 
	\left(\OO_\Delta^\dagger\right){}^{\alpha_1\ldots\alpha_{\bar\ell}}_{\dot\beta_1\ldots\dot\beta_{\ell}}(x)
	\left(\bar s^{\dot\beta_1}\ldots \bar s^{\dot\beta_\ell}\right)
	\left(s_{\alpha_1}\ldots s_{\alpha_{\bar\ell}}\right)\\
	&= (-1)^{\ell+\bar\ell}
	\left(\OO_\Delta^\dagger\right){}_{\alpha_1\ldots\alpha_{\bar\ell}}^{\dot\beta_1\ldots\dot\beta_{\ell}}(x)
	\left(\bar s_{\dot\beta_1}\ldots \bar s_{\dot\beta_\ell}\right)
	\left(s^{\alpha_1}\ldots s^{\alpha_{\bar\ell}}\right).
\end{aligned}
\end{equation}
In the second equality we used the epsilon symbols to raise and lower all indices. 
The factor $(-1)^{\ell+\bar\ell}$ arises from these manipulations.

It is convenient to use the following definition
\begin{equation}
	\left(\overline\OO_\Delta\right){}_{\alpha_1\ldots\alpha_{\bar\ell}}^{\dot\beta_1\ldots\dot\beta_{\ell}}(x) \equiv
	(-1)^{\ell+\bar\ell}
	\left(\OO_\Delta^\dagger\right){}_{\alpha_1\ldots\alpha_{\bar\ell}}^{\dot\beta_1\ldots\dot\beta_{\ell}}(x).
\end{equation}
Note that this definition applies only to local operators and not to the spinor polarizations $s$ and $\bar s$. 
Using this notation, we can write
\begin{equation}
		\left(\OO_\Delta^{(\ell, \bar \ell)}(x,s,\bar s) \right)^\dagger = 	\overline\OO_\Delta^{(\bar\ell, \ell)}(x,s,\bar s).
\end{equation}

\paragraph{Two-point functions}
Two-point functions of generic operators are completely fixed by conformal invariance up to an overall normalisation. According to \cite{Cuomo:2017wme} they read
\begin{equation}
	\label{eq:2pt_normalization}
	\< \overline \OO_{\Delta}^{(\bar\ell, \ell)}(\point_1) \OO_{\Delta}^{(\ell, \bar \ell)}(\point_2) \> = i^{\ell-\bar\ell}
	\mathcal{N}_{\OO}\times \frac{\left(\hat\II^{12}\right)^\ell\left(\hat\II^{21}\right)^{\bar\ell}}{\left(x_{12}^2\right)^{\Delta+\frac{\ell+\bar\ell}{2}}},
\end{equation}
where the two operators should have the same scaling dimensions and 
\begin{equation}
	\label{eq:invariant_I}
	x_{ij}\equiv |x_i^\mu - x_j^\mu|,\qquad
	\hat\II^{ij}\equiv x_{ij}^\mu (\bar s_i\bar\sigma_\mu s_j).
\end{equation}
The phase factor $i^{\ell-\bar\ell}$ is required by unitarity. 
The real non-negative constant $\mathcal{N}_{\OO}>0$ defines the normalization of the operators, which we choose as
\begin{equation}
	\label{eq:norm_convention}
	\mathcal{N}_{\OO}=1.
\end{equation}
Notice that non-zero two-point functions only exist when one operator is the hermitian conjugate of the second one as in \eqref{eq:2pt_normalization}. The normalization choice in \eqref{eq:norm_convention} is adopted for all local operators, with the exception of the conserved current $J(\point)$ and the stress-tensor $T(\point)$, whose normalizations are instead fixed by the global $U(1)$ symmetry and the conformal algebra respectively.

\paragraph{Central charges}
The two-point functions of the conserved current and the stress-tensor are completely fixed by conformal invariance and the corresponding algebraic structures. 
We have
\begin{equation}
	\label{eq:JJ_TT}
	\< J(\point_1) J(\point_2) \> = C_J\times \frac{\hat\II^{12}\hat\II^{21}}{\left(x_{12}^2\right)^{4}},\qquad
	\< T(\point_1) T(\point_2) \> = C_T\times \frac{\left(\hat\II^{12}\hat\II^{21}\right)^2}{\left(x_{12}^2\right)^{6}},
\end{equation}
where $C_J$ and $C_T$ are referred to as the current and stress-tensor central charges, respectively. 
It is convenient to introduce the rescaled operators
\begin{equation}
	\label{eq:hat_J}
	\widehat J(\point) \equiv C_J^{-1/2} J(\point),\qquad
	\widehat T(\point) \equiv C_T^{-1/2} T(\point),
\end{equation}
which are then normalized according to the general convention \eqref{eq:2pt_normalization}, namely
\begin{equation}
	\label{eq:conserved_rescaled_2pt}
	\<\widehat J(\point_1) \widehat J(\point_2) \> = \frac{\hat\II^{12}\hat\II^{21}}{\left(x_{12}^2\right)^{4}},\qquad
	\< \widehat T(\point_1) \widehat T(\point_2) \> = \frac{\left(\hat\II^{12}\right)^2\left(\hat\II^{21}\right)^2}{\left(x_{12}^2\right)^{6}}.
\end{equation}
We can therefore treat $\widehat J$ and $\widehat T$ as special cases of generic operators,
\begin{equation}
	\cO^{(1,1)}_{\Delta=3}(\point) = \widehat J(\point),\qquad
	\cO^{(2,2)}_{\Delta=4}(\point) = \widehat T(\point).
\end{equation}

\paragraph{Unitarity bound}
For an operator in the $(\ell,\bar \ell)$ Lorentz representation there is a lower-bound on its scaling dimension $\Delta$. This bound is called the unitarity bound and it reads
\be
\label{eq:unitarity_bound}
\De\geq\De_\text{unitary}(\ell,\bar\ell)=&
\begin{cases}
	2+\frac{\ell+\bar\ell}{2}, & \ell\bar \ell\neq 0,\\
	1+\frac{\ell+\bar\ell}{2},& \ell\bar\ell=0.
\end{cases}
\ee
Operators which saturate the first line are conserved. Operators which saturate the second line are free.

\paragraph{Operator labels}
Given an operator in the $(\ell, \bar \ell)$ Lorentz representation, it is convenient to define the following number
\begin{equation}
	p \equiv |\ell-\bar \ell|.
\end{equation}
The number $p\geq 0$ is a non-negative integer.
The operators with $p=0$ are called traceless symmetric (TS). The operators with $p\neq 0$ are non-traceless symmetric (NTS). For NTS operators will adopt the following naming conventions: the operators with $\ell\leq \bar\ell$ are called primal and the operators with $\ell\geq \bar\ell$ are called dual.
It is convenient to write explicitly the labels primal and dual, namely
\begin{equation}
	\label{eq:primal_and_dual}
	\OO{}_{\Delta,\, \text{primal}}^{(\ell,\,\ell+p)}(\point)
	\qquad\text{and}\qquad
	\OO{}_{\Delta,\, \text{dual}}^{(\ell+p,\,\ell)}(\point).
\end{equation}

\paragraph{Neutral operators}
In this paper we restrict our analysis to the neutral subsector of the CFT, consisting of operators with vanishing $U(1)$ charge. 
Consequently, Hermitian conjugation acts on such operators in a simple manner. 
For $p=0$ traceless symmetric operators, we can choose a basis in which
\begin{equation}
	\label{eq:hermicity_p=0}
	\OO_\Delta^{(\ell,\ell)}(\point) = \overline\OO_\Delta^{(\ell,\ell)}(\point).
\end{equation}
For non-traceless symmetric operators with $p \neq 0$, Hermitian conjugation acts non-trivially only on the Lorentz indices. 
In this case, we can always choose the operator basis such that Hermitian conjugation relates the primal and dual operators defined in \eqref{eq:primal_and_dual}, namely
\begin{equation}
	\label{eq:relation_primal_dual}
	\OO_{\Delta,\, \text{primal}}^{(\ell,\,\ell+p)}(\point) =  \overline\OO^{(\ell,\, \ell+p)}_{\Delta,\,\text{dual}}(\point).
\end{equation}


\section{Three-point functions}
\label{app:three-point_functions}
Let us consider the three-point correlation function of two conserved currents and a generic operator $\cO$ with scaling dimension $\Delta$ and Lorentz representation $(\ell,\bar\ell)$. 
This correlator can be written as
\begin{equation}
	\label{eq:3pt_general}
	\<J(\point_1)J(\point_2)\OO_\Delta^{(\ell,\bar\ell)}(\point_3)\> 
	= \sum_{a=1}^{N} \lambda_{JJ\cO,\, a}\, \mathbf{T}_{JJ\cO, \,a} (\point_1, \point_2, \point_3),
\end{equation}
where $\lambda_{JJ\cO, a}$ are the OPE coefficients, $\mathbf{T}_{JJ\cO,a}$ denote the tensor structures, and $N$ is the number of linearly independent conformally invariant tensor structures. 

Throughout, we assume that all operators are space-like separated, so the left-hand side of \eqref{eq:3pt_general} is symmetric under the interchange of the two currents. 
It also satisfies the current conservation conditions at points $\point_1$ and $\point_2$. The right-hand side of \eqref{eq:3pt_general} must obey the same constraints. 
In this section we present the basis of tensor structures $\mathbf{T}_{JJ\cO,a}$ used in this work; this basis is constructed to automatically satisfy both conservation and permutation symmetry.

The three-point function \eqref{eq:3pt_general} is non-vanishing only when the third operator has $p = 0$, $2$, or $4$. 
In what follows, we discuss each of these three cases separately. 
A summary of the corresponding numbers of tensor structures in \eqref{eq:3pt_general} is provided in table \ref{tab:structures}.

For traceless-symmetric operators, each tensor structure is either parity even or parity odd. 
It is therefore convenient to assign explicit labels to both the OPE coefficients and the tensor structures:
\begin{equation}
	p=0:\qquad
	\lambda^{0,\,\ell,\,\pm}_{JJ\cO}\; \mathbf{T}^{0,\,\ell,\,\pm}_{JJ\cO}.
\end{equation}
For traceless-symmetric operators, these labels indicate the values of $p$, $\ell$, and the parity ($+$ or $-$). 
For non-traceless-symmetric operators, instead of the parity label, we introduce the primal and dual labels, and use the notation
\begin{equation}
	p\neq0:\qquad
	\lambda^{p,\,\ell,\,\text{primal}}_{JJ\cO}\; \mathbf{T}^{p,\,\ell,\,\text{primal}}_{JJ\cO}
	\qquad\text{and}\qquad
	\lambda^{p,\,\ell,\,\text{dual}}_{JJ\cO}\; \mathbf{T}^{p,\,\ell,\,\text{dual}}_{JJ\cO}.
\end{equation}

\begin{table}[h]
	\centering
	\begin{tabular}{ |c | c | c| }
		\hline
		$p$ & Spin $\ell$ & Number of structures \\ 
		\hline\hline
		\multirow{3}{*}{$0$} 
		& $\ell = 0$ & $1^{+}$ \\ 
		& $\ell \geq 1$ (odd) & $1^{-}$ \\ 
		& $\ell \geq 2$ (even) & $2^{+}$ \\ 
		\hline
		\multirow{3}{*}{$2$} 
		& $\ell = 0$ & $0$ \\ 
		& $\ell \geq 1$ (odd) & $1^\text{primal} + 1^\text{dual}$ \\ 
		& $\ell \geq 2$ (even) & $1^\text{primal} + 1^\text{dual}$ \\ 
		\hline
		\multirow{2}{*}{$4$} 
		& $\ell \geq 0$ (even) & $1^\text{primal} + 1^\text{dual}$ \\ 
		& $\ell \geq 1$ (odd) & $0$ \\ 
		\hline
	\end{tabular}
	\caption{Number of conserved and permutation-invariant tensor structures for different values of $p$ and $\ell$. 
		For traceless-symmetric operators, the tensor structures are classified as parity even (denoted by the superscript $+$) or parity odd (denoted by $-$).}
	\label{tab:structures}
\end{table}

\subsection{Basis of tensor structures}
The basis of three-point tensor structures is provided in the ancillary file \verb|data_JJO.m| accompanying this paper. 
The syntax and structure of this file are described in \cite{Karateev:2026xxx}. 
Below we explicitly list the corresponding basis used in this work. 
The file \verb|data_JJO.m| also includes the differential basis required for constructing conformal blocks, which are discussed in the next appendix. 
We do not provide additional details on the differential basis here.

The kinematic factor is defined as
\begin{equation}
	\label{eq:kinamtic_factor}
	\mathcal{K}_3^{(\ell,\bar\ell)}\equiv
	\left(x_{12}^2\right)^{\frac{\kappa}{2}-4}
	\left(x_{23}^2\right)^{-\frac{\kappa}{2}}
	\left(x_{13}^2\right)^{-\frac{\kappa}{2}},\qquad 
	\kappa\equiv\Delta+\frac{\ell+\bar\ell}{2},
\end{equation}
where $\mathcal{K}_3^{(\ell,\bar\ell)}$ captures the overall coordinate dependence of three-point functions. 
In the formulas below, we will make frequent use of the tensor invariants $\hat\II^{ij}$, $\hat\JJ^{k}_{ij}$, $\hat\KK^{ij}_{k}$, and $\hat{\overline \KK}^{ij}_{k}$, whose definitions can be found in appendix~D of \cite{Cuomo:2017wme}. 
The invariant $\hat\II^{ij}$ was already defined in \eqref{eq:invariant_I}, while $\hat\JJ^{k}_{ij}$ is given by
\begin{equation}
	\label{eq:invariant_J}
	\hat\JJ^{k}_{ij}= \,\frac{x_{ik}^2x_{jk}^2}{x_{ij}^{2}}\,
	\times\bigg( \frac{x_{ik}^\mu}{x_{ik}^2}-\frac{x_{jk}^\mu}{x_{jk}^2} \bigg)\times\,
	(\bar s_k \bar \sigma_\mu  s_k).
\end{equation}
The invariants $\hat\KK^{ij}_{k}$ and $\hat{\overline \KK}^{ij}_{k}$ have slightly more involved expressions, which we do not reproduce here.

\paragraph{$p=0$ operators} The $p=0$ three-point functions read as
\begin{align}
	\nn
	\ell=0:\quad
	\<J(\point_1)J(\point_2)\OO_\Delta^{(0,0))}(\point_3)\> &= \lambda^{0,0,+}_{JJ\OO}\; \mathbf{T}^{0,\,0,\,+}_{JJ\cO},\\
	\label{eq:casep0}
	\ell\geq 1\text{ (odd)}:\quad
	\<J(\point_1)J(\point_2)\OO_\Delta^{(\ell,\ell))}(\point_3)\> &=  \lambda^{\ell,0,-}_{JJ\OO}\; \mathbf{T}^{0,\,\ell,\,-}_{JJ\cO},\\
	\nn
	\ell\geq 2\text{ (even)}:\quad
	\<J(\point_1)J(\point_2)\OO_\Delta^{(\ell,\ell))}(\point_3)\> &= \lambda^{\ell,0,+}_{JJ\OO,\,1}\; \mathbf{T}^{0,\,\ell,\,+}_{JJ\cO,1} + \lambda^{\ell,0,+}_{JJ\OO,\,2}\;\mathbf{T}^{0,\,\ell,\,+}_{JJ\cO,2},
\end{align}
where the tensor structure appearing in the first two lines read as
\begin{align}
	\mathbf{T}^{0,\,0,\,+}_{JJ\cO} &\equiv
	\mathcal{K}_3^{(0,0)}\;
	\left(\hat\JJ^1_{23}\hat\JJ^2_{13}-2\,(1-3/\Delta)\hat\II^{12}\hat\II^{21}\right),\\
	\mathbf{T}^{0,\,\ell,\,-}_{JJ\cO}&\equiv
	 \mathcal{K}_3^{(\ell,\ell)}
	\left(\hat\II^{12}\hat\II^{23}\hat\II^{31}+\hat\II^{21}\hat\II^{13}\hat\II^{32}\right)
	\left(\hat\JJ^3_{12}\right)^{\ell-1}.
\end{align}
In the \verb|data_JJO.m| file we treat $\ell=1$ and $\ell\geq 3$ parity odd tensor structures differently because of the differences in their differential bases. The tensor structures appearing in the last line of \eqref{eq:casep0} are given by
\begin{multline}
	 \mathbf{T}^{0,\,\ell,\,+}_{JJ\cO,1}  \equiv  \frac{1}{\Delta^{2}}\,\mathcal{K}_3^{(\ell,\ell)}\left(\hat\JJ^3_{12}\right)^{\ell-2}
	 \bigg(
	 4\ell (\Delta-3)\hat\II^{13}\hat\II^{23}\hat\II^{31}\hat\II^{32}\\
	 +2(4+\ell-\Delta)(\Delta-3)\hat\II^{12}\hat\II^{21}\left(\hat\JJ^3_{12}\right)^{2}
	 -(4+\ell-\Delta)(\ell+\Delta)\hat\JJ^1_{23}\hat\JJ^2_{13}\left(\hat\JJ^3_{12}\right)^{2}
	 \bigg),
\end{multline}
together with
\begin{multline}
	\mathbf{T}^{0,\,\ell,\,+}_{JJ\cO,2}  \equiv  \frac{1}{\Delta\,\ell}\,\mathcal{K}_3^{(\ell,\ell)}\left(\hat\JJ^3_{12}\right)^{\ell-2}
	\bigg(
	8\ell (\Delta-3)\hat\II^{13}\hat\II^{23}\hat\II^{31}\hat\II^{32}
	+4(2+\ell)(\Delta-3)\hat\II^{12}\hat\II^{21}\left(\hat\JJ^3_{12}\right)^{2}\\
	-4(\Delta+\ell\Delta-2\ell)\hat\JJ^1_{23}\hat\JJ^2_{13}\left(\hat\JJ^3_{12}\right)^{2}
	-4\ell(\Delta-3)
	\left(
	\hat\II^{13}\hat\II^{31}\hat\JJ^2_{13}-
	\hat\II^{23}\hat\II^{32}\hat\JJ^1_{23}
	\right)\hat\JJ^3_{12}
	\bigg).
\end{multline}

\paragraph{$p=2$ operators} 
Let us now discuss the case of $p=2$ operators. 
When $p=2$ and $\ell=0$, the three-point function \eqref{eq:3pt_general} is non-zero only if $\Delta=2$, as required by current conservation and permutation symmetry. 
The $p=2$ operator with $\Delta=2$ saturates the unitarity bound and therefore corresponds to a free operator, see \cite{Weinberg:2012cd}.

For the primal operator $\cO$ we have
\begin{equation}
	\<J(\point_1)J(\point_2)\OO_\Delta^{(\ell,\ell+2)}(\point_3)\> = \lambda^{2,\,\ell,\,\text{primal}}_{JJ\OO}
	\;\mathbf{T}^{2,\,\ell,\,\text{primal}}_{JJ\cO},
\end{equation}
where the explicit form of the tensor structure depends on whether $\ell$ is even or odd. The tensor structures are given by
\begin{multline}
	\label{eq:T2odd_primal}
    \ell=\text{odd}:\quad\mathbf{T}^{2,\,\ell,\,\text{primal}}_{JJ\cO} 
	\equiv \\
	\frac{i}{\Delta}\,
	\mathcal{K}_3^{(\ell,\ell+2)}
	\left(\hat\JJ^3_{12}\right)^{\ell}
	\Bigg(
	(\Delta-\ell-6) \left(
	\hat\II^{12}\hat\II^{31}\hat{\bar\KK}^{23}_{1}-\hat\II^{21}\hat\II^{32}\hat{\bar\KK}^{13}_{2}
	\right)
	+2(\ell+2)
	\hat\II^{31}\hat\II^{32}\hat{\bar\KK}^{12}_{3}\Bigg),
\end{multline}
\begin{multline}
	\label{eq:T2even_primal}
	\ell=\text{even}:\quad\mathbf{T}^{2,\,\ell,\,\text{primal}}_{JJ\cO} 
	\equiv 
	\frac{i}{2\Delta}\,
	\mathcal{K}_3^{(\ell,\ell+2)}
	\left(\hat\JJ^3_{12}\right)^{\ell-1}
	\Bigg(
	(\Delta-\ell-6) \left(
	\hat\II^{12}\hat\II^{31}\hat{\bar\KK}^{23}_{1}+\hat\II^{21}\hat\II^{32}\hat{\bar\KK}^{13}_{2}
	\right)\hat\JJ^3_{12}\\
	+2(\Delta-2)
	\left(
	2\hat\II^{32}\hat\II^{23}\hat\II^{31}\hat{\bar\KK}^{13}_{2}-\hat\II^{32}\hat\II^{31}\hat{\bar\KK}^{12}_{3}\hat\JJ^3_{12}
	\right)
	\Bigg).
\end{multline}

For the dual operator $\cO$ we have
\begin{equation}
	\<J(\point_1)J(\point_2)\OO_\Delta^{(\ell+2,\ell)}(\point_3)\> = \lambda^{2,\,\ell,\,\text{dual}}_{JJ\OO}
	\;\mathbf{T}^{2,\,\ell,\,\text{dual}}_{JJ\cO}.
\end{equation}
As in the primal case, the explicit form of the tensor structure depends on whether $\ell$ is even or odd. The tensor structures are given by
\begin{multline}
	\label{eq:T2odd_dual}
	\ell=\text{odd}:\quad\mathbf{T}^{2,\,\ell,\,\text{dual}}_{JJ\cO} 
	\equiv \\
	\frac{i}{\Delta}\,
	\mathcal{K}_3^{(\ell+2,\ell)}
	\left(\hat\JJ^3_{12}\right)^{\ell}
	\Bigg(
	(\Delta-\ell-6) \left(
	\hat\II^{21}\hat\II^{13}\hat{\KK}^{23}_{1}-\hat\II^{12}\hat\II^{23}\hat{\KK}^{13}_{2}
	\right)
	+2(\ell+2)
	\hat\II^{13}\hat\II^{23}\hat{\KK}^{12}_{3}\Bigg),
\end{multline}
\begin{multline}
	\label{eq:T2even_dual}
	\ell=\text{even}:\quad\mathbf{T}^{2,\,\ell,\,\text{dual}}_{JJ\cO} 
	\equiv 
	\frac{i}{2\Delta}\,
	\mathcal{K}_3^{(\ell+2,\ell)}
	\left(\hat\JJ^3_{12}\right)^{\ell-1}
	\Bigg(
	(\Delta-\ell-6) \left(
	\hat\II^{21}\hat\II^{13}\hat{\KK}^{23}_{1}+\hat\II^{12}\hat\II^{23}\hat{\KK}^{13}_{2}
	\right)\hat\JJ^3_{12}\\
	-2(\Delta-2)
	\left(
	2\hat\II^{32}\hat\II^{23}\hat\II^{13}\hat{\KK}^{13}_{2}+\hat\II^{23}\hat\II^{13}\hat{\KK}^{12}_{3}\hat\JJ^3_{12}
	\right)
	\Bigg).
\end{multline}

\paragraph{$p=4$ operators} 
\label{sec:p=4}
For $p=4$ operators the three-point function \eqref{eq:3pt_general} is non-zero only if $\ell$ is even. The three-point functions of primal and dual operators in this case read
\begin{align}
	\label{eq:p=4_structure_primal}
	\<J(\point_1)J(\point_2)\cO^{(\ell,\ell+4)}_{\De}(\point_3)\>&\equiv
	\lambda^{4,\,\ell,\,\text{primal}}_{JJ\OO}\;\mathbf{T}^{4,\,\ell,\,\text{primal}}_{JJ\cO},\\
	\label{eq:p=4_structure_dual}
	\<J(\point_1)J(\point_2)\cO^{(\ell+4,\ell)}_{\De}(\point_3)\>&\equiv
	\lambda^{4,\,\ell,\,\text{dual}}_{JJ\OO}\;	\mathbf{T}^{4,\,\ell,\,\text{dual}}_{JJ\cO}
\end{align}
where the basis of tensor structures is chosen to be
\begin{align}
	\label{eq:P4even_primal}
	\mathbf{T}^{4,\,\ell,\,\text{primal}}_{JJ\cO} &= \mathcal{K}_3^{(\ell,\ell+4)}\;\hat\II^{31}\hat\II^{32}\hat{\bar\KK}^{13}_{2}\hat{\bar\KK}^{23}_{1}
	\left(\hat\JJ^3_{12}\right)^\ell,\\
	\label{eq:P4even_dual}
	\mathbf{T}^{4,\,\ell,\,\text{dual}}_{JJ\cO} &= \mathcal{K}_3^{(\ell+4,\ell)}\;\hat\II^{13}\hat\II^{23}\hat\KK^{13}_{2}\hat\KK^{23}_{1}
	\left(\hat\JJ^3_{12}\right)^\ell.
\end{align}

\subsection{Relations between the OPE coefficients}
\label{app:relations_OPE_coefficients}
Consider the following simple relation
\begin{equation}
	\label{eq:condition}
	\begin{aligned}
		\<J(\point_1)J(\point_2)\OO(\point_3)\>^* &= 
		\<\overline\OO(\point_3)J(\point_1)J(\point_2)\>\\
		&= 
		\<J(\point_1)J(\point_2)\overline\OO(\point_3)\>,
	\end{aligned}
\end{equation}
where the last line holds only for space-like separated operators only. Let us now derive relations on the OPE coefficients, which follow from \eqref{eq:condition}.

\paragraph{$p=0$ operators}
We start from traceless symmetric operators. Recall that these operators are hermitian \eqref{eq:hermicity_p=0}. For them the condition \eqref{eq:condition} can be rewritten as
\begin{equation}
	\<J(\point_1)J(\point_2)\OO(\point_3)\>^* = \<J(\point_1)J(\point_2)\OO(\point_3)\>.
\end{equation}
Let us now substitute the decomposition \eqref{eq:casep0}. 
The transformation properties of tensor invariants under complex conjugation are given in equations (D.9) and (D.10) of \cite{Cuomo:2017wme}, and hold under the assumption \eqref{eq:relation_s_sbar}. 
From these relations, we conclude that the basis of tensor structures chosen in the $p=0$ case satisfies
\begin{equation}
	\left(\mathbf{T}^{0,\,0,\,+}_{JJ\cO}\right)^* = \mathbf{T}^{0,\,0,\,+}_{JJ\cO},\qquad
	\left( \mathbf{T}^{0,\,\ell,\,-}_{JJ\cO}\right)^*=- \mathbf{T}^{0,\,\ell,\,-}_{JJ\cO},\qquad
	\left(\mathbf{T}^{0,\,\ell,\,+}_{JJ\cO,a}\right)^* = \mathbf{T}^{0,\,\ell,\,+}_{JJ\cO,a}.
\end{equation}
Using these relations, we immediately obtain
\begin{equation}
	\label{eq:TS_OPE_properties}
	\left(\lambda^{0,0,+}_{JJ\OO}\right)^* = \lambda^{0,0,+}_{JJ\OO},\qquad
	\left(\lambda^{\ell,0,-}_{JJ\OO}\right)^* = -\lambda^{\ell,0,-}_{JJ\OO},\qquad
	\left(\lambda^{\ell,0,+}_{JJ\OO,\,a}\right)^* = \lambda^{\ell,0,+}_{JJ\OO,\,a}.
\end{equation}
In other words, in the chosen basis of three-point tensor structures, the $p=0$ OPE coefficients are purely real for even $\ell$ and purely imaginary for odd $\ell$.

\paragraph{$p\neq 0$ operators}
Non-traceless symmetric operators are related under hermitian conjugation as in \eqref{eq:relation_primal_dual}. Using this relation, we can rewrite \eqref{eq:condition} as
\begin{equation}
	\<J(\point_1)J(\point_2)\OO_\text{primal}(\point_3)\>^* = \<J(\point_1)J(\point_2)\OO_\text{dual}(\point_3)\>.
\end{equation}
In the previous subsection we have chosen the basis of three-point tensor structures for $p\neq 0$ operators such that
\begin{equation}
	\label{eq:relation_OPE_NTS}
	\left(\mathbf{T}^{p,\,\ell,\,\text{primal}}_{JJ\cO}\right)^* = 	\mathbf{T}^{p,\,\ell,\,\text{dual}}_{JJ\cO}.
\end{equation}
Consequently, the corresponding OPE coefficients satisfy
\begin{equation}
	\label{eq:primal_dual_OPE_relation}
	\left(\lambda^{p,\,\ell,\,\text{primal}}_{JJ\cO}\right)^* = \lambda^{p,\,\ell,\,\text{dual}}_{JJ\cO}.
\end{equation}


\section{Conformal blocks}
\label{app:conformal_blocks}

In this section, we decompose the following four-point function into conformal blocks:
\begin{equation}
	\label{eq:JJJJ}
	\<J(\point_1) J(\point_2) J(\point_3) J(\point_4)\>.
\end{equation}
This is achieved by inserting the identity operator,
\begin{equation}
	\label{eq:identity}
	1 = \sum_\alpha |\alpha\>\<\alpha| = \sum_\cO \cO(\pp_0)|0\>\bowtie\<0|\overline\cO(\pp_0),
\end{equation}
into the middle of the correlator \eqref{eq:JJJJ}.  
In \eqref{eq:identity}, the label $\alpha$ runs over all states in the theory, which are created by primary operators and their descendants. In the second equality, the sum is restricted to primary operators $\cO$, while the contribution of their descendants is encoded in the $\bowtie$ operation. This operation was introduced in section 4.1 of \cite{Karateev:2017jgd} as a double integral involving the two-point function $\<\overline\cO\cO\>$. We will not need its explicit form here.  

Using these relations, we can rewrite the four-point function as
\begin{equation}
	\label{eq:four-point_function_decomposition}
	\begin{aligned}
		\< J(\pp_1) J(\pp_2) J(\pp_3) J(\pp_4)\> &= \sum_\cO
		\< J(\pp_1) J(\pp_2)\cO(\pp_0)\>\bowtie\<\overline\cO(\pp_0) J(\pp_3) J(\pp_4)\>\\
		&= \sum_\cO
		\< J(\pp_1) J(\pp_2)\cO(\pp_0)\>\bowtie\< J(\pp_4) J(\pp_3)\overline\cO(\pp_0)\>.
	\end{aligned}
\end{equation}
In the second line, we used the fact that bosonic operators commute at space-like separation.  
We refer to the three-point function appearing to the left of the $\bowtie$ symbol as the \emph{left three-point function}, and to the one on the right as the \emph{right three-point function}.

\paragraph{Operator content}
According to the discussion at the end of appendix \ref{app:notation}, the only non-zero contributions to \eqref{eq:four-point_function_decomposition} arise from $p=0$ traceless symmetric (TS) operators and from $p=2,4$ primal and dual non-traceless symmetric (NTS) operators. 
We can schematically denote these contributions as
\begin{equation}
	\cO = 
	\left\{
	\OO{}_\text{TS},\;
	\OO{}_\text{primal},\;
	\OO{}_\text{dual}
	\right\}.
\end{equation}
Let us now write explicitly the corresponding contributions in \eqref{eq:four-point_function_decomposition}. 
Using \eqref{eq:hermicity_p=0} and \eqref{eq:relation_primal_dual}, we obtain
\begin{equation}
	\begin{aligned}
		\< J(\pp_1) J(\pp_2) J(\pp_3) J(\pp_4)\> 
		&= \sum_{\cO_\text{TS}} \< J(\pp_1) J(\pp_2)\cO_\text{TS}(\pp_0)\>\bowtie\< J(\pp_4) J(\pp_3)\cO_\text{TS}(\pp_0)\>\\
		&+ \sum_{\cO_\text{primal}} \< J(\pp_1) J(\pp_2)\cO_\text{primal}(\pp_0)\>\bowtie\< J(\pp_4) J(\pp_3)\cO_\text{dual}(\pp_0)\>\\
		&+ \sum_{\cO_\text{dual}} \< J(\pp_1) J(\pp_2)\cO_\text{dual}(\pp_0)\>\bowtie\< J(\pp_4) J(\pp_3)\cO_\text{primal}(\pp_0)\>.
	\end{aligned}
\end{equation}
The operation $\bowtie$ is symmetric under the exchange of the left and right three-point functions. 
Consequently, the last line above can be equivalently written as
\begin{multline}
	\< J(\pp_1) J(\pp_2)\cO_\text{dual}(\pp_0)\>\bowtie\< J(\pp_4) J(\pp_3)\cO_\text{primal}(\pp_0)\> = \\
	\< J(\pp_4) J(\pp_3)\cO_\text{primal}(\pp_0)\>\bowtie \< J(\pp_1) J(\pp_2)\cO_\text{dual}(\pp_0)\>.
\end{multline}

We now decompose each three-point function in terms of the tensor structures defined in appendix \ref{app:three-point_functions}. 
This yields
\begin{multline}
	\label{eq:conformal_block_decomposition_preliminary}
	\< J(\pp_1) J(\pp_2) J(\pp_3) J(\pp_4)\> = \sum_{\cO_\text{TS}}
	\sum_{a,b}
	\lambda_{JJ\cO,\, a}  \lambda_{JJ\cO,\, b} G_{\cO,ab} (\point_1, \point_2, \point_3, \point_4)\\
	+\sum_{\cO_\text{primal}}
	\lambda^{\text{primal}}_{JJ\cO} \lambda^{\text{dual}}_{JJ\cO} \left(
	G_{\cO}^\text{primal} (\point_1, \point_2, \point_3, \point_4)
	+G_{\cO}^\text{primal} (\point_3, \point_4, \point_1, \point_2) \right),
\end{multline}
where $\lambda^{\text{dual}}_{JJ\cO} = \left(\lambda^{\text{primal}}_{JJ\cO}\right)^*$ according to \eqref{eq:primal_dual_OPE_relation}.
The conformal blocks corresponding to TS operators are defined as
\begin{equation}
	G^{ab}_{\cO}(\pp_1,\pp_2,\pp_3,\pp_4) \equiv 	\mathbf{T}_{JJ\cO,\,a}(\point_1,\point_2,\point_0)\bowtie\mathbf{T}_{JJ\cO,\,b}(\point_4,\point_3,\point_0),
\end{equation}
where the three-point tensor structures $\mathbf{T}_{JJ\cO,\,a}$ are those given in \eqref{eq:casep0}. 
For the primal NTS blocks, we similarly define
\begin{align}
	G_{\cO}^\text{primal}(\pp_1,\pp_2,\pp_3,\pp_4) \equiv
	\mathbf{T}^{\text{primal}}_{JJ\cO}(\point_1,\point_2,\point_0)\bowtie\mathbf{T}^{\text{dual}}_{JJ\cO}(\point_4,\point_3,\point_0),
\end{align}
where the primal and dual three-point tensor structures are defined in \eqref{eq:T2odd_primal}, \eqref{eq:T2even_primal}, \eqref{eq:T2odd_dual}, and \eqref{eq:T2even_dual} for $p=2$ operators, and in \eqref{eq:P4even_primal} and \eqref{eq:P4even_dual} for $p=4$ operators.

\paragraph{Explicit expressions}
Let us now present the conformal block decomposition \eqref{eq:conformal_block_decomposition_preliminary} in a more explicit form. 
First, we write explicitly the contribution of the identity operator, which is a special case of the TS operator. 
Second, we separate the TS contributions into the cases $\ell=0$, $\ell=1,3,5,\ldots$, and $\ell=2,4,6,\ldots$. 
Third, we make use of the relations between the OPE coefficients derived in section \ref{app:three-point_functions}. 
Finally, we define each conformal block precisely in terms of the three-point tensor structures introduced in section \ref{app:three-point_functions}. 
The resulting expression reads
\begin{multline}
	\label{eq:conformal_block_decomposition}
	\< J(\pp_1) J(\pp_2) J(\pp_3) J(\pp_4)\> =\< J(\pp_1) J(\pp_2)\> \<J(\pp_3) J(\pp_4)\> \\
	+\text{contributionTS}
	+\text{contributionP2Primal}
	+\text{contributionP4Primal}.
\end{multline}
Below we will write explicitly the three contributions appearing in this decomposition. 
Before doing so, however, let us recall the discussion preceding equation \eqref{eq:3pt_rescaled_example}. In this work we consider correlation functions that are appropriately rescaled by powers of $C_J$. In particular, as explained above \eqref{eq:3pt_rescaled_example}, it is convenient to work with the rescaled current $\widehat J$ defined in \eqref{eq:hat_J}. Accordingly, we must work in practice with the four-point function
\begin{equation}
	\langle \widehat J(\pp_1) \widehat J(\pp_2) \widehat J(\pp_3) \widehat J(\pp_4) \rangle
	=
	C_J^{-2}\langle J(\pp_1) J(\pp_2) J(\pp_3) J(\pp_4) \rangle.
\end{equation}
Substituting the conformal block decomposition \eqref{eq:conformal_block_decomposition} into this expression, we obtain
\begin{multline}
	\label{eq:conformal_block_decomposition_rescaled}
\< \widehat J(\pp_1) \widehat J(\pp_2) \widehat J(\pp_3) \widehat J(\pp_4)\> =
	 \frac{\hat\II^{12}\hat\II^{21}}{\left(x_{12}^2\right)^{4}} \frac{\hat\II^{34}\hat\II^{43}}{\left(x_{34}^2\right)^{4}}\\
	+C_J^{-2}\left(\text{contributionTS}
	+\text{contributionP2Primal}
	+\text{contributionP4Primal}\right),
\end{multline}
where in order to write the first term in the right-hand side we have used the results \eqref{eq:hat_J} and \eqref{eq:conserved_rescaled_2pt}.

\subparagraph{Traceless-symmetric contribution}
The contribution from the TS operators is given by the following expression
\begin{multline}
	\text{contributionTS} \equiv 
	\sum_{\Delta} \p{\lambda^{0,0,+}_{JJ\cO}}^2
	G^{0,0,+}_{\Delta} (\point_1, \point_2, \point_3, \point_4)
	+ \sum_{\Delta} \sum_{\ell=1,3,5,\ldots} \p{\lambda^{0,\ell,-}_{JJ\cO}}^2
	G^{0,\ell,-}_{\Delta} (\point_1, \point_2, \point_3, \point_4)\\
	+ \sum_{\Delta}\sum_{\ell=2,4,6,\ldots}
	\sum_{a,b}
	\lambda^{0,\ell,+}_{JJ\cO,\, a} \lambda^{0,\ell,+}_{JJ\cO,\, b}
	G^{0,\ell,+}_{\Delta,ab} (\point_1, \point_2, \point_3, \point_4).
\end{multline}
Using reality properties of OPE coefficients, we can exhibit the positive contributions (note the minus signs for parity-odd exchanges),
\begin{multline}
	\label{eq:contributionTS}
	\text{contributionTS} \equiv 
	\sum_{\Delta} \left| \lambda^{0,0,+}_{JJ\cO} \right|^2
	G^{0,0,+}_{\Delta} (\point_1, \point_2, \point_3, \point_4)
	- \sum_{\Delta} \sum_{\ell=1,3,5,\ldots} \left|\lambda^{0,\ell,-}_{JJ\cO} \right|^2
	G^{0,\ell,-}_{\Delta} (\point_1, \point_2, \point_3, \point_4)\\
	+ \sum_{\Delta}\sum_{\ell=2,4,6,\ldots}
	\sum_{a,b}
	\lambda^{0,\ell,+}_{JJ\cO,\, a} \left(\lambda^{0,\ell,+}_{JJ\cO,\, b}\right)^*
	G^{0,\ell,+}_{\Delta,ab} (\point_1, \point_2, \point_3, \point_4).
\end{multline}
The conformal blocks appearing here are defined as
\begin{align}
	\label{eq:CB_spin=0}
	G^{0,0,+}_{\Delta} (\point_1, \point_2, \point_3, \point_4) &\equiv \mathbf{T}^{0,\,0,\,+}_{JJ\cO}(\point_1, \point_2, \point_0) \bowtie \mathbf{T}^{0,\,0,\,+}_{JJ\cO}(\point_4, \point_3, \point_0),\\
	\label{eq:CB_spin=odd}
	G^{0,\ell,-}_{\Delta} (\point_1, \point_2, \point_3, \point_4) &\equiv \mathbf{T}^{0,\,\ell,\,-}_{JJ\cO}(\point_1, \point_2, \point_0) \bowtie \mathbf{T}^{0,\,\ell,\,-}_{JJ\cO}(\point_4, \point_3, \point_0),\\
	\label{eq:CB_spin=even}
	G^{0,\ell,+}_{\Delta,ab} (\point_1, \point_2, \point_3, \point_4) &\equiv \mathbf{T}^{0,\,\ell,\,+}_{JJ\cO,a}(\point_1, \point_2, \point_0) \bowtie \mathbf{T}^{0,\,\ell,\,+}_{JJ\cO,b}(\point_4, \point_3, \point_0).
\end{align}
The first conformal block can be found in the \verb|block_P0Spin0.m| file. The second conformal block can be found in the \verb|block_P0SpinOdd.m| file. There are four conformal blocks in the last line. They correspond to different values of $(a,b)=\{(1,1),\; (1,2),\; (2,1),\; (2,2)\}$. They can be found in the \verb|block_P0SpinEven_11.m|, \verb|block_P0SpinEven_12.m|, \verb|block_P0SpinEven_21.m| and \verb|block_P0SpinEven_22.m| files.

In writing the expression \eqref{eq:contributionTS}, we implicitly assumed that all exchanged operators $\cO$ are normalised in exactly the same way dictated by equation \eqref{eq:2pt_normalization}. 
In particular, when $\Delta=3$ and $\ell=1$, the exchanged operator is the rescaled conserved current $\widehat J$. 
To account for this fact in \eqref{eq:contributionTS}, we must effectively perform the replacement
\begin{equation}
	\Delta=3,\quad
	\ell=1:\qquad
	\left|\lambda^{0,\ell,-}_{JJ\cO} \right|^2 \longrightarrow C_J^{-1} \left|\lambda^{0,1,-}_{JJJ} \right|^2.
\end{equation}
Let us now recall the presence of the additional factor $C_J^{-2}$ in \eqref{eq:conformal_block_decomposition_rescaled}. 
As a result, the overall factor multiplying the conformal block \eqref{eq:CB_spin=odd} appearing in \eqref{eq:contributionTS} due to the exchange of the rescaled current $\widehat J$ is given by
\begin{equation}
	\label{eq:coefficient_JJJ_block}
	\frac{\left|\lambda^{0,1,-}_{JJ\cO} \right|^2}{C_J^3}.
\end{equation}
This explains why, in our setup, we can only constrain the quantity $P_J$, first introduced in \eqref{eq:P_J} and given explicitly by \eqref{eq:coefficient_JJJ_block}, rather than the ’t~Hooft anomaly by itself.

Analogously, when $\Delta=4$ and $\ell=2$, the exchanged operator is the rescaled stress tensor. 
To incorporate this into \eqref{eq:contributionTS}, we must perform the replacement
\begin{equation}
	\Delta=4,\quad
	\ell=2:\qquad
	\lambda^{0,\ell,+}_{JJ\cO,\, a} \left(\lambda^{0,\ell,+}_{JJ\cO,\, b}\right)^*
	\longrightarrow
	C_T^{-1} \lambda^{0,2,+}_{JJT,\, a} \left(\lambda^{0,2,+}_{JJT,\, b}\right)^*.
\end{equation}
Once again recalling the additional factor $C_J^{-2}$ in \eqref{eq:conformal_block_decomposition_rescaled}, we find that the coefficient multiplying the conformal block \eqref{eq:CB_spin=even} appearing in \eqref{eq:conformal_block_decomposition_rescaled}  due to the exchange of the rescaled stress-tensor $\widehat T$ is given by
\begin{equation}
	\label{eq:coefficient_JJT_block}
	\frac{\lambda^{0,2,+}_{JJT,\, a} \left(\lambda^{0,2,+}_{JJT,\, b}\right)^*}{C_J^2 C_T}.
\end{equation}
Thus, in our setup we are able to constrain the quantity $P_T$, first introduced in \eqref{eq:P_T} and given explicitly by \eqref{eq:coefficient_JJT_block}. 
Finally, note that the OPE coefficient $\lambda^{0,2,+}_{JJT,\, b}$ is real according to \eqref{eq:TS_OPE_properties}. 
As a result, \eqref{eq:coefficient_JJT_block} is identical to \eqref{eq:P_T}.

\subparagraph{Non traceless-symmetric contribution}
The $p=2$ contribution takes the form
\begin{equation}
	\text{contributionP2Primal} \equiv
	\sum_{\Delta} \sum_{\ell\geq 1}
	\left|\lambda^{2,\,\ell,\,\text{primal}}_{JJ\cO}\right|^2 \left(1+\pi_{13}\pi_{24} \right)G^{2,\ell,\text{primal}}_{\Delta} (\point_1, \point_2, \point_3, \point_4).
\end{equation}
The corresponding conformal block is defined as
\begin{equation}
	G^{2,\ell,\text{primal}}_{\Delta} (\point_1, \point_2, \point_3, \point_4) \equiv \mathbf{T}^{2,\ell,\text{primal}}_{JJ\cO}(\point_1, \point_2, \point_0) \bowtie \mathbf{T}^{2,\ell,\text{dual}}_{JJ\cO}(\point_4, \point_3, \point_0),
\end{equation}
where $\pi_{ij}$ denotes the operator that permutes the points $\point_i$ and $\point_j$ in the conformal block. 
The structure of the conformal block depends on whether $\ell$ is odd or even. 
The block with $p=2$ and odd $\ell$ is provided in the file \verb|block_P2PrimalSpinOdd.m|, 
while that with $p=2$ and even $\ell$ can be found in \verb|block_P2PrimalSpinEven.m|.

Finally, the contribution of the $p=4$ blocks is given by
\begin{multline}
	\text{contributionP4Primal} \equiv\\
	\sum_{\Delta} \sum_{\ell=0,2,4,\ldots}
	\left|\lambda^{4,\,\ell,\,\text{primal}}_{JJ\cO}\right|^2 \left(1+\pi_{13}\pi_{24} \right)G^{4,\ell,\text{primal}}_{\Delta} (\point_1, \point_2, \point_3, \point_4).
\end{multline}
The corresponding conformal block is defined as
\begin{equation}
	G^{4,\ell,\text{primal}}_{\Delta} (\point_1, \point_2, \point_3, \point_4) \equiv \mathbf{T}^{4,\ell,\text{primal}}_{JJ\cO}(\point_1, \point_2, \point_0) \bowtie \mathbf{T}^{4,\ell,\text{dual}}_{JJ\cO}(\point_4, \point_3, \point_0).
\end{equation}
The conformal blocks with $p=4$ and even $\ell$ can be found in the file \verb|block_P4PrimalSpinEven.m|.

\paragraph{Computation of the blocks}
All conformal blocks are computed automatically from the three-point tensor structure data provided in the \verb|data_JJO.m| file, using the automated framework developed in \cite{Karateev:2026xxx}, which builds upon the already existing technology.

\section{Crossing equations}
\label{app:crossing_equations}

Let us now discuss the tensor structures of the four-point correlator $\< JJJJ \>$. 
This correlator admits 70 independent tensor structures in total: 43 parity-even and 27 parity-odd. 
We can therefore write
\begin{equation}
	\label{eq:4-point_funtion}
	\<J(\point_1)J(\point_2)J(\point_3)J(\point_4)\> = \sum_{i=1}^{70} g_i(z, \bar z) \mathbf{T}_{JJJJ,\, i}(\point_1, \point_2, \point_3, \point_4),
\end{equation}
where $g_i(z, \bar z)$ are functions of the two conformally invariant cross-ratios
\begin{equation}
	z \bar z = \frac{x_{12}^2x_{34}^2}{x_{13}^2x_{24}^2},\qquad
	(1-z)(1-\bar z) = \frac{x_{14}^2x_{23}^2}{x_{13}^2x_{24}^2}.
\end{equation}

The basis of tensor structures is constructed unambiguously in the conformal frame (CF). 
Following the notation of section 4 of \cite{Cuomo:2017wme}, we write
\begin{equation}
	\label{eq:structure_CF}
	\mathbf{T}_{JJJJ,\, i}(\point_1^\text{CF}, \point_2^\text{CF}, \point_3^\text{CF}, \point_4^\text{CF}) = \structgeneral.
\end{equation}
All 70 tensor structures are obtained by taking all possible combinations of 
$q_1, q_2, q_3, q_4 = \pm \tfrac{1}{2}$ and 
$\bar q_1, \bar q_2, \bar q_3, \bar q_4 = \pm \tfrac{1}{2}$, 
subject to the constraint
\begin{equation}
	q_1+q_2+q_3+q_4-\bar q_1-\bar q_2-\bar q_3-\bar q_4 = 0.
\end{equation}

Suppose a given tensor structure \eqref{eq:structure_CF} is multiplied by some function $f(z, \bar z)$. 
In the \texttt{CFTs4D} notation, such a term is denoted by
\begin{equation}
	\text{CF4pt}\Big[\{3, 3, 3, 3\}, \{1, 1, 1, 1\}, \{1, 1, 1, 1\}, \{q_1, q_2, q_3, q_4\}, 
	\{\bar q_1, \bar q_2, \bar q_3, \bar q_4\}, f[z,\bar z]\Big].
\end{equation}
By comparing \eqref{eq:4-point_funtion} with the conformal block decomposition \eqref{eq:conformal_block_decomposition}, 
we can express the functions $g_i(z, \bar z)$ in terms of conformal block components and the OPE coefficients.

\paragraph{Crossing equations}
For space-like separated points, the four-point function satisfies
\begin{equation}
	\<J(\point_1)J(\point_2)J(\point_3)J(\point_4)\> = \<J(\point_1)J(\point_4)J(\point_3)J(\point_2)\>.
\end{equation}
Using the decomposition \eqref{eq:4-point_funtion} on both sides of this equality, we find
\begin{equation}
	\sum_{i=1}^{70} \Big(g_i(z, \bar z) \mathbf{T}_{JJJJ,\, i}(\point_1, \point_2, \point_3, \point_4) -
	g_i(1-z, 1-\bar z) \mathbf{T}_{JJJJ,\, i}(\point_1, \point_4, \point_3, \point_2) \Big) = 0.
\end{equation}
We can express the crossed tensor structures in terms of the original basis as
\begin{equation}
	\mathbf{T}_{JJJJ,\, i}(\point_1, \point_4, \point_3, \point_2) = \sum_{j=1}^{70}M_{ij}\,	\mathbf{T}_{JJJJ,\, j}(\point_1, \point_2, \point_3, \point_4),
\end{equation}
where $M_{ij}$ is the \emph{crossing matrix}. 
Substituting this relation into the above equation gives
\begin{equation}
	\label{eq:crossing_equations_full}
	\sum_{j=1}^{70} \Big(g_j(z, \bar z) -
	\sum_{i=1}^{70}g_i(1-z, 1-\bar z)M_{ij} \Big) \mathbf{T}_{JJJJ,\, j}(\point_1, \point_2, \point_3, \point_4)   = 0.
\end{equation}
If all tensor structures were linearly independent, we would have obtained 70 scalar crossing equations,
\begin{equation}
	g_j(z, \bar z) - g_i(1-z, 1-\bar z)M_{ij} = 0.
\end{equation}
However, most of these equations are linearly dependent. 
Only seven of them are linearly independent.

\paragraph{Independent crossing equations}
We now focus on the independent crossing equations. 
First, we assume parity invariance, which leaves 43 parity-even structures. 
Furthermore, imposing the $(12)(34)$, $(13)(24)$, and $(14)(23)$ permutation symmetries reduces the number of independent structures to 19. 
Current conservation imposes additional constraints, and one finds that only seven two-variable degrees of freedom remain, without any line or point degrees of freedom. The analysis of the independent degrees of freedom is fully analogous to~\cite{Dymarsky:2017yzx}.

We extract the corresponding seven crossing equations by applying seven independent linear functionals to the full equation \eqref{eq:crossing_equations_full}. 
These functionals are provided in the file \verb|unconstrained_JJJJ_structures.m|.

As an example, the first functional in our file reads
\begin{equation}
	\text{functional}_1 = \frac{1}{16} \text{coefficientOf} \struct{-\frac{1}{2}}{-\frac{1}{2}}{-\frac{1}{2}}{-\frac{1}{2}}{-\frac{1}{2}}{-\frac{1}{2}}{-\frac{1}{2}}{-\frac{1}{2}}+  \frac{1}{16} \text{coefficientOf} \struct{+\frac{1}{2}}{+\frac{1}{2}}{+\frac{1}{2}}{+\frac{1}{2}}{+\frac{1}{2}}{+\frac{1}{2}}{+\frac{1}{2}}{+\frac{1}{2}}.
\end{equation}
This functional instructs one to extract the coefficients of the two corresponding tensor structures, divide each by 16, and then add them together.

\section{The numerical setup}
\label{app:numerical_setup}
The numerical setup was implemented using hyperion \cite{hyperion}. 
The conformal blocks were computed using \cite{blocks4d_gitlab, blocks4d_haskell_gitlab}, 
and we employed the rational approximation described in \cite{Chang:2025mwt}. 
All parameters were chosen such that truncation and approximation errors are well under control.

We denote by $\textbf{param}$ the collective set of parameters used in the numerical analysis, 
summarized in tables \ref{tab:params}, \ref{tab:parms-nmax-dep}, and \ref{tab:sdpb-params}. 
To estimate numerical uncertainties, we compared results obtained with a given parameter choice $\textbf{param}$ 
to those computed using a more precise reference setup $\textbf{param}^{*}$. 
In particular, for representative central charge maximizations, we verified that the relative deviation satisfies
\begin{equation}
	\frac{C_T^{\textbf{param}}-C_T^{\textbf{param}^{*}}}{C_T^{\textbf{param}^*}} < 10^{-6}.
\end{equation}
The corresponding parameter choices are summarized in the tables below.

\begin{table}[h]
	\centering
	\begin{tabular}{c|c|c}
		parameter & $\textbf{param}$ & $\textbf{param}^{*}$ \\
		\hline
		{\tt twistGap} & $10^{-6}$ & $10^{-6}$ \\
		\hline
		{\tt order \cite{blocks4d_gitlab}} & 115 & 125 \\
		\hline
		{\tt keptPoleOrder \cite{blocks4d_gitlab}} & 115 & 125 \\
		\hline
		{\tt keptPoleOrder\_final \cite {blocks4d_gitlab}} & 115 & 125 \\
		\hline
		$n_\mathrm{poles}$ \cite{Chang:2025mwt} & 30 & 40 \\
		\hline
		{\tt precision} \cite{hyperion} & 1152 & 1280 \\
		\hline
		{\tt floatFormat} \cite{hyperion} & 5 & 10 \\
	\end{tabular}
	\caption{The parameter choices used in the numerics. The first column shows the values used to create the plots. The second shows a set of more precise reference values against which we compared to estimate the errors due to the various truncation/approximation choices. To improve numerical stability, positivity conditions that demand positivity above the unitarity bound were replaced with $\Delta\geq \De_\text{unitary}(\ell,\bar\ell) + \text{twistGap}$.}\label{tab:params}
\end{table}

\begin{table}[h]
	\centering
	\begin{tabular}{c|c}
		spins &  \\
		\hline
		$\textbf{param}$ &
		$\{0,\ldots,C\}\ \cup\ \{C+1,C+2,C+5,C+6,C+9,C+10,\ldots, U\}$ \\
		$\textbf{param}^{*}$&$\{0,\ldots,U\}$
	\end{tabular}
	\caption{The spins included in the positivity conditions. We include all spins up to $C=33+2\,{\tt nmax}$ and then every other consecutive pair up to $U=33+4\,{\tt nmax}+9$. The first row shows the values used to create the plots. The second shows a larger set of spins used to estimate the error due to the spin truncation.}\label{tab:parms-nmax-dep}
\end{table}

\begin{table}
	\centering
	\begin{tabular}{c|c|c}
		parameter & feasibility & maximization \\
		\hline
		{\tt precision} & 1152 & 1152 \\
		\hline
		{\tt findPrimalFeasible} & {\tt False} & {\tt False} \\
		\hline
		{\tt findDualFeasible} & {\tt False} & {\tt False} \\
		\hline
		{\tt detectPrimalFeasibleJump} & {\tt True} & {\tt False} \\
		\hline
		{\tt detectDualFeasibleJump} & {\tt True} & {\tt False} \\
		\hline
		{\tt primalErrorThreshold} & $10^{-30}$ & $10^{-30}$ \\
		\hline
		{\tt dualErrorThreshold} & $10^{-30}$ & $10^{-30}$ \\
		\hline
		{\tt dualityGapThreshold} & $10^{-30}$ & $10^{-30}$ \\
		\hline
		{\tt initialMatrixScalePrimal} & $10^{40}$ & $10^{40}$ \\
		\hline
		{\tt initialMatrixScaleDual} & $10^{40}$ & $10^{40}$ \\
		\hline
		{\tt feasibleCenteringParameter} & $0.1$ & $0.1$ \\
		\hline
		{\tt infeasibleCenteringParameter} & $0.3$ & $0.3$ \\
		\hline
		{\tt stepLengthReduction} & $0.7$ & $0.7$ \\
		\hline
		{\tt maxComplementarity} & $10^{100}$ & $10^{100}$ \\
		\hline
	\end{tabular}
	\caption{SDPB parameters used for feasibility and maximization runs. These parameters were kept fixed for both the $\textbf{param}$ and $\textbf{param}^{*}$ parameter choices.}
	\label{tab:sdpb-params}
\end{table}

\FloatBarrier
\bibliographystyle{JHEP}
\bibliography{refs}

@article{Rychkov:2023wsd,
    author = "Rychkov, Slava and Su, Ning",
    title = "{New developments in the numerical conformal bootstrap}",
    eprint = "2311.15844",
    archivePrefix = "arXiv",
    primaryClass = "hep-th",
    doi = "10.1103/RevModPhys.96.045004",
    journal = "Rev. Mod. Phys.",
    volume = "96",
    number = "4",
    pages = "045004",
    year = "2024"
}

@article{Li:2017ddj,
    author = "Li, Daliang and Meltzer, David and Stergiou, Andreas",
    title = "{Bootstrapping mixed correlators in 4D $ \mathcal{N} $ = 1 SCFTs}",
    eprint = "1702.00404",
    archivePrefix = "arXiv",
    primaryClass = "hep-th",
    reportNumber = "CERN-TH-2017-024",
    doi = "10.1007/JHEP07(2017)029",
    journal = "JHEP",
    volume = "07",
    pages = "029",
    year = "2017"
}

@article{Manenti:2019jds,
    author = "Manenti, Andrea",
    title = "{Differential operators for superconformal correlation functions}",
    eprint = "1910.12869",
    archivePrefix = "arXiv",
    primaryClass = "hep-th",
    doi = "10.1007/JHEP04(2020)145",
    journal = "JHEP",
    volume = "04",
    pages = "145",
    year = "2020"
}

@article{Manenti:2019kbl,
    author = "Manenti, Andrea and Stergiou, Andreas and Vichi, Alessandro",
    title = "{Implications of ANEC for SCFTs in four dimensions}",
    eprint = "1905.09293",
    archivePrefix = "arXiv",
    primaryClass = "hep-th",
    doi = "10.1007/JHEP01(2020)093",
    journal = "JHEP",
    volume = "01",
    pages = "093",
    year = "2020"
}

@article{tHooft:1979rat,
    author = "'t Hooft, Gerard",
    editor = "'t Hooft, Gerard and Itzykson, C. and Jaffe, A. and Lehmann, H. and Mitter, P. K. and Singer, I. M. and Stora, R.",
    title = "{Naturalness, chiral symmetry, and spontaneous chiral symmetry breaking}",
    reportNumber = "PRINT-80-0083 (UTRECHT)",
    doi = "10.1007/978-1-4684-7571-5_9",
    journal = "NATO Sci. Ser. B",
    volume = "59",
    pages = "135--157",
    year = "1980"
}

@article{Rattazzi:2008pe,
	Archiveprefix = {arXiv},
	Author = {Rattazzi, Riccardo and Rychkov, Vyacheslav S. and Tonni, Erik and Vichi, Alessandro},
	Doi = {10.1088/1126-6708/2008/12/031},
	Eprint = {0807.0004},
	Journal = {JHEP},
	Pages = {031},
	Primaryclass = {hep-th},
	Slaccitation = {%%CITATION = 0807.0004;%%},
	Title = {{Bounding scalar operator dimensions in 4D CFT}},
	Volume = {12},
	Year = {2008}}

@article{Rychkov:2009ij,
	Archiveprefix = {arXiv},
	Author = {Rychkov, Vyacheslav S. and Vichi, Alessandro},
	Doi = {10.1103/PhysRevD.80.045006},
	Eprint = {0905.2211},
	Journal = {Phys. Rev.},
	Pages = {045006},
	Primaryclass = {hep-th},
	Slaccitation = {%%CITATION = 0905.2211;%%},
	Title = {{Universal Constraints on Conformal Operator Dimensions}},
	Volume = {D80},
	Year = {2009}}

@article{Poland:2010wg,
	Archiveprefix = {arXiv},
	Author = {Poland, David and Simmons-Duffin, David},
	Doi = {10.1007/JHEP05(2011)017},
	Eprint = {1009.2087},
	Journal = {JHEP},
	Pages = {017},
	Primaryclass = {hep-th},
	Slaccitation = {%%CITATION = ARXIV:1009.2087;%%},
	Title = {{Bounds on 4D Conformal and Superconformal Field Theories}},
	Volume = {1105},
	Year = {2011}}

@article{Rattazzi:2010gj,
	Archiveprefix = {arXiv},
	Author = {Rattazzi, Riccardo and Rychkov, Slava and Vichi, Alessandro},
	Doi = {10.1103/PhysRevD.83.046011},
	Eprint = {1009.2725},
	Journal = {Phys. Rev.},
	Pages = {046011},
	Primaryclass = {hep-th},
	Slaccitation = {%%CITATION = 1009.2725;%%},
	Title = {{Central Charge Bounds in 4D Conformal Field Theory}},
	Volume = {D83},
	Year = {2011}}

@article{Rattazzi:2010yc,
	Archiveprefix = {arXiv},
	Author = {Rattazzi, Riccardo and Rychkov, Slava and Vichi, Alessandro},
	Doi = {10.1088/1751-8113/44/3/035402},
	Eprint = {1009.5985},
	Journal = {J. Phys.},
	Pages = {035402},
	Primaryclass = {hep-th},
	Slaccitation = {%%CITATION = 1009.5985;%%},
	Title = {{Bounds in 4D Conformal Field Theories with Global Symmetry}},
	Volume = {A44},
	Year = {2011}}

@article{Vichi:2011ux,
	Archiveprefix = {arXiv},
	Author = {Vichi, Alessandro},
	Doi = {10.1007/JHEP01(2012)162},
	Eprint = {1106.4037},
	Journal = {JHEP},
	Pages = {162},
	Primaryclass = {hep-th},
	Slaccitation = {%%CITATION = ARXIV:1106.4037;%%},
	Title = {{Improved bounds for CFT's with global symmetries}},
	Volume = {1201},
	Year = {2012}}

@article{Poland:2011ey,
	Archiveprefix = {arXiv},
	Author = {Poland, David and Simmons-Duffin, David and Vichi, Alessandro},
	Doi = {10.1007/JHEP05(2012)110},
	Eprint = {1109.5176},
	Journal = {JHEP},
	Pages = {110},
	Primaryclass = {hep-th},
	Slaccitation = {%%CITATION = ARXIV:1109.5176;%%},
	Title = {{Carving Out the Space of 4D CFTs}},
	Volume = {1205},
	Year = {2012}}

@article{Mack:1975je,
	Author = {Mack, G.},
	Doi = {10.1007/BF01613145},
	Journal = {Commun.Math.Phys.},
	Pages = {1},
	Reportnumber = {DESY 75/50},
	Slaccitation = {%%CITATION = CMPHA,55,1;%%},
	Title = {{All Unitary Ray Representations of the Conformal Group $SU(2,2)$ with Positive Energy}},
	Volume = {55},
	Year = {1977}}

@article{SimmonsDuffin:2012uy,
	Archiveprefix = {arXiv},
	Author = {Simmons-Duffin, David},
	Doi = {10.1007/JHEP04(2014)146},
	Eprint = {1204.3894},
	Journal = {JHEP},
	Pages = {146},
	Primaryclass = {hep-th},
	Slaccitation = {%%CITATION = ARXIV:1204.3894;%%},
	Title = {{Projectors, Shadows, and Conformal Blocks}},
	Volume = {1404},
	Year = {2014}}

@article{Costa:2011dw,
	Archiveprefix = {arXiv},
	Author = {Costa, Miguel S. and Penedones, Joao and Poland, David and Rychkov, Slava},
	Doi = {10.1007/JHEP11(2011)154},
	Eprint = {1109.6321},
	Journal = {JHEP},
	Pages = {154},
	Primaryclass = {hep-th},
	Reportnumber = {LPTENS-11-37},
	Slaccitation = {%%CITATION = ARXIV:1109.6321;%%},
	Title = {{Spinning Conformal Blocks}},
	Volume = {1111},
	Year = {2011}}

@article{Costa:2011mg,
	Archiveprefix = {arXiv},
	Author = {Costa, Miguel S. and Penedones, Joao and Poland, David and Rychkov, Slava},
	Doi = {10.1007/JHEP11(2011)071},
	Eprint = {1107.3554},
	Journal = {JHEP},
	Pages = {071},
	Primaryclass = {hep-th},
	Reportnumber = {LPTENS-11-22, NSF-KITP-11-128},
	Slaccitation = {%%CITATION = ARXIV:1107.3554;%%},
	Title = {{Spinning Conformal Correlators}},
	Volume = {1111},
	Year = {2011}}

@article{Kos:2013tga,
	Archiveprefix = {arXiv},
	Author = {Kos, Filip and Poland, David and Simmons-Duffin, David},
	Doi = {10.1007/JHEP06(2014)091},
	Eprint = {1307.6856},
	Journal = {JHEP},
	Pages = {091},
	Slaccitation = {%%CITATION = ARXIV:1307.6856;%%},
	Title = {{Bootstrapping the $O(N)$ vector models}},
	Volume = {1406},
	Year = {2014}}

@article{Kos:2015mba,
      author         = "Kos, Filip and Poland, David and Simmons-Duffin, David
                        and Vichi, Alessandro",
      title          = "{Bootstrapping the O(N) Archipelago}",
      journal        = "JHEP",
      volume         = "11",
      year           = "2015",
      pages          = "106",
      doi            = "10.1007/JHEP11(2015)106",
      eprint         = "1504.07997",
      archivePrefix  = "arXiv",
      primaryClass   = "hep-th",
      reportNumber   = "CERN-PH-TH-2015-097",
      SLACcitation   = "%%CITATION = ARXIV:1504.07997;%%"
}

@article{El-Showk:2014dwa,
	Archiveprefix = {arXiv},
	Author = {El-Showk, Sheer and Paulos, Miguel F. and Poland, David and Rychkov, Slava and Simmons-Duffin, David and Vichi, Alessandro},
	Doi = {10.1007/s10955-014-1042-7},
	Eprint = {1403.4545},
	Journal = {J.Stat.Phys.},
	Month = {June},
	Pages = {869},
	Primaryclass = {hep-th},
	Reportnumber = {CERN-PH-TH-2014-038, NSF-KITP-14-022},
	Slaccitation = {%%CITATION = ARXIV:1403.4545;%%},
	Title = {{Solving the 3d Ising Model with the Conformal Bootstrap II. $c$-Minimization and Precise Critical Exponents}},
	Volume = {157},
	Year = {2014}}

@article{Kos:2016ysd,
      author         = "Kos, Filip and Poland, David and Simmons-Duffin, David
                        and Vichi, Alessandro",
      title          = "{Precision Islands in the Ising and $O(N)$ Models}",
      journal        = "JHEP",
      volume         = "08",
      year           = "2016",
      pages          = "036",
      doi            = "10.1007/JHEP08(2016)036",
      eprint         = "1603.04436",
      archivePrefix  = "arXiv",
      primaryClass   = "hep-th",
      reportNumber   = "CERN-TH-2016-050",
      SLACcitation   = "%%CITATION = ARXIV:1603.04436;%%"
}

@article{Caracciolo:2014cxa,
      author         = "Caracciolo, Francesco and Echeverri, Alejandro Castedo
                        and von Harling, Benedict and Serone, Marco",
      title          = "{Bounds on OPE Coefficients in 4D Conformal Field
                        Theories}",
      journal        = "JHEP",
      volume         = "10",
      year           = "2014",
      pages          = "20",
      doi            = "10.1007/JHEP10(2014)020",
      eprint         = "1406.7845",
      archivePrefix  = "arXiv",
      primaryClass   = "hep-th",
      SLACcitation   = "%%CITATION = ARXIV:1406.7845;%%"
}

@article{Elkhidir:2014woa,
	Archiveprefix = {arXiv},
	Author = {Elkhidir, Emtinan and Karateev, Denis and Serone, Marco},
	Doi = {10.1007/JHEP01(2015)133},
	Eprint = {1412.1796},
	Journal = {JHEP},
	Pages = {133},
	Primaryclass = {hep-th},
	Reportnumber = {SISSA-65-2014-FISI},
	Slaccitation = {%%CITATION = ARXIV:1412.1796;%%},
	Title = {{General Three-Point Functions in 4D CFT}},
	Volume = {1501},
	Year = {2015}}

@article{Simmons-Duffin:2015qma,
      author         = "Simmons-Duffin, David",
      title          = "{A Semidefinite Program Solver for the Conformal
                        Bootstrap}",
      journal        = "JHEP",
      volume         = "06",
      year           = "2015",
      pages          = "174",
      doi            = "10.1007/JHEP06(2015)174",
      eprint         = "1502.02033",
      archivePrefix  = "arXiv",
      primaryClass   = "hep-th",
      SLACcitation   = "%%CITATION = ARXIV:1502.02033;%%"
}

@article{Iliesiu:2015qra,
      author         = "Iliesiu, Luca and Kos, Filip and Poland, David and Pufu,
                        Silviu S. and Simmons-Duffin, David and Yacoby, Ran",
      title          = "{Bootstrapping 3D Fermions}",
      journal        = "JHEP",
      volume         = "03",
      year           = "2016",
      pages          = "120",
      doi            = "10.1007/JHEP03(2016)120",
      eprint         = "1508.00012",
      archivePrefix  = "arXiv",
      primaryClass   = "hep-th",
      SLACcitation   = "%%CITATION = ARXIV:1508.00012;%%"
}

@article{Osborn:1993cr,
      author         = "Osborn, H. and Petkou, A. C.",
      title          = "{Implications of conformal invariance in field theories
                        for general dimensions}",
      journal        = "Annals Phys.",
      volume         = "231",
      year           = "1994",
      pages          = "311-362",
      doi            = "10.1006/aphy.1994.1045",
      eprint         = "hep-th/9307010",
      archivePrefix  = "arXiv",
      primaryClass   = "hep-th",
      reportNumber   = "DAMTP-93-31",
      SLACcitation   = "%%CITATION = HEP-TH/9307010;%%"
}

@article{Dolan:2000ut,
      author         = "Dolan, F. A. and Osborn, H.",
      title          = "{Conformal four point functions and the operator product
                        expansion}",
      journal        = "Nucl. Phys.",
      volume         = "B599",
      year           = "2001",
      pages          = "459-496",
      doi            = "10.1016/S0550-3213(01)00013-X",
      eprint         = "hep-th/0011040",
      archivePrefix  = "arXiv",
      primaryClass   = "hep-th",
      reportNumber   = "DAMTP-2000-125",
      SLACcitation   = "%%CITATION = HEP-TH/0011040;%%"
}

@article{Chester:2016wrc,
      author         = "Chester, Shai M. and Pufu, Silviu S.",
      title          = "{Towards bootstrapping QED$_{3}$}",
      journal        = "JHEP",
      volume         = "08",
      year           = "2016",
      pages          = "019",
      doi            = "10.1007/JHEP08(2016)019",
      eprint         = "1601.03476",
      archivePrefix  = "arXiv",
      primaryClass   = "hep-th",
      reportNumber   = "PUPT-2494",
      SLACcitation   = "%%CITATION = ARXIV:1601.03476;%%"
}

@article{Iliesiu:2017nrv,
      author         = "Iliesiu, Luca and Kos, Filip and Poland, David and Pufu,
                        Silviu S. and Simmons-Duffin, David",
      title          = "{Bootstrapping 3D Fermions with Global Symmetries}",
      journal        = "JHEP",
      volume         = "01",
      year           = "2018",
      pages          = "036",
      doi            = "10.1007/JHEP01(2018)036",
      eprint         = "1705.03484",
      archivePrefix  = "arXiv",
      primaryClass   = "hep-th",
      reportNumber   = "PUPT-2524",
      SLACcitation   = "%%CITATION = ARXIV:1705.03484;%%"
}

@article{Hofman:2008ar,
      author         = "Hofman, Diego M. and Maldacena, Juan",
      title          = "{Conformal collider physics: Energy and charge
                        correlations}",
      journal        = "JHEP",
      volume         = "05",
      year           = "2008",
      pages          = "012",
      doi            = "10.1088/1126-6708/2008/05/012",
      eprint         = "0803.1467",
      archivePrefix  = "arXiv",
      primaryClass   = "hep-th",
      SLACcitation   = "%%CITATION = ARXIV:0803.1467;%%"
}

@article{Dymarsky:2017yzx,
      author         = "Dymarsky, Anatoly and Kos, Filip and Kravchuk, Petr and
                        Poland, David and Simmons-Duffin, David",
      title          = "{The 3d Stress-Tensor Bootstrap}",
      journal        = "JHEP",
      volume         = "02",
      year           = "2018",
      pages          = "164",
      doi            = "10.1007/JHEP02(2018)164",
      eprint         = "1708.05718",
      archivePrefix  = "arXiv",
      primaryClass   = "hep-th",
      reportNumber   = "CALT-TH-2017-043",
      SLACcitation   = "%%CITATION = ARXIV:1708.05718;%%"
}

@article{Elkhidir:2017iov,
      author         = "Elkhidir, Emtinan and Karateev, Denis",
      title          = "{Scalar-Fermion Analytic Bootstrap in 4D}",
      year           = "2017",
      eprint         = "1712.01554",
      archivePrefix  = "arXiv",
      primaryClass   = "hep-th",
      SLACcitation   = "%%CITATION = ARXIV:1712.01554;%%"
}

@article{Karateev:2017jgd,
      author         = "Karateev, Denis and Kravchuk, Petr and Simmons-Duffin,
                        David",
      title          = "{Weight Shifting Operators and Conformal Blocks}",
      journal        = "JHEP",
      volume         = "02",
      year           = "2018",
      pages          = "081",
      doi            = "10.1007/JHEP02(2018)081",
      eprint         = "1706.07813",
      archivePrefix  = "arXiv",
      primaryClass   = "hep-th",
      reportNumber   = "CALT-TH-2017-031",
      SLACcitation   = "%%CITATION = ARXIV:1706.07813;%%"
}

@article{Cuomo:2017wme,
      author         = "Cuomo, Gabriel Francisco and Karateev, Denis and
                        Kravchuk, Petr",
      title          = "{General Bootstrap Equations in 4D CFTs}",
      journal        = "JHEP",
      volume         = "01",
      year           = "2018",
      pages          = "130",
      doi            = "10.1007/JHEP01(2018)130",
      eprint         = "1705.05401",
      archivePrefix  = "arXiv",
      primaryClass   = "hep-th",
      reportNumber   = "CALT-TH-2017-23, SISSA-23-2017-FISI",
      SLACcitation   = "%%CITATION = ARXIV:1705.05401;%%"
}

@article{Costa:2016hju,
      author         = "Costa, Miguel S. and Hansen, Tobias and Penedones, Jo�o
                        and Trevisani, Emilio",
      title          = "{Projectors and seed conformal blocks for traceless
                        mixed-symmetry tensors}",
      journal        = "JHEP",
      volume         = "07",
      year           = "2016",
      pages          = "018",
      doi            = "10.1007/JHEP07(2016)018",
      eprint         = "1603.05551",
      archivePrefix  = "arXiv",
      primaryClass   = "hep-th",
      reportNumber   = "CERN-TH-2016-080",
      SLACcitation   = "%%CITATION = ARXIV:1603.05551;%%"
}

@article{Cordova:2017dhq,
      author         = "Cordova, Clay and Diab, Kenan",
      title          = "{Universal Bounds on Operator Dimensions from the Average
                        Null Energy Condition}",
      journal        = "JHEP",
      volume         = "02",
      year           = "2018",
      pages          = "131",
      doi            = "10.1007/JHEP02(2018)131",
      eprint         = "1712.01089",
      archivePrefix  = "arXiv",
      primaryClass   = "hep-th",
      SLACcitation   = "%%CITATION = ARXIV:1712.01089;%%"
}

@article{Manenti:2018xns,
      author         = "Manenti, Andrea and Stergiou, Andreas and Vichi,
                        Alessandro",
      title          = "{R-current three-point functions in 4d $\mathcal{N}=1$
                        superconformal theories}",
      year           = "2018",
      eprint         = "1804.09717",
      archivePrefix  = "arXiv",
      primaryClass   = "hep-th",
      reportNumber   = "CERN-TH-2018-096, CERN-TH-2018-096",
      SLACcitation   = "%%CITATION = ARXIV:1804.09717;%%"
}

@article{Chester:2017vdh,
      author         = "Chester, Shai M. and Iliesiu, Luca V. and Mezei, Mark and
                        Pufu, Silviu S.",
      title          = "{Monopole Operators in $U(1)$ Chern-Simons-Matter
                        Theories}",
      year           = "2017",
      eprint         = "1710.00654",
      archivePrefix  = "arXiv",
      primaryClass   = "hep-th",
      reportNumber   = "PUPT-2537",
      SLACcitation   = "%%CITATION = ARXIV:1710.00654;%%"
}

@article{DeGrand:2015zxa,
      author         = "DeGrand, Thomas",
      title          = "{Lattice tests of beyond Standard Model dynamics}",
      journal        = "Rev. Mod. Phys.",
      volume         = "88",
      year           = "2016",
      pages          = "015001",
      doi            = "10.1103/RevModPhys.88.015001",
      eprint         = "1510.05018",
      archivePrefix  = "arXiv",
      primaryClass   = "hep-ph",
      reportNumber   = "COLO-HEP-589",
      SLACcitation   = "%%CITATION = ARXIV:1510.05018;%%"
}

@article{Dolan:2011dv,
      author         = "Dolan, F. A. and Osborn, H.",
      title          = "{Conformal Partial Waves: Further Mathematical Results}",
      year           = "2011",
      eprint         = "1108.6194",
      archivePrefix  = "arXiv",
      primaryClass   = "hep-th",
      reportNumber   = "DAMTP-11-64, CCTP-2011-32",
      SLACcitation   = "%%CITATION = ARXIV:1108.6194;%%"
}

@article{Dolan:2003hv,
      author         = "Dolan, F. A. and Osborn, H.",
      title          = "{Conformal partial waves and the operator product
                        expansion}",
      journal        = "Nucl. Phys.",
      volume         = "B678",
      year           = "2004",
      pages          = "491-507",
      doi            = "10.1016/j.nuclphysb.2003.11.016",
      eprint         = "hep-th/0309180",
      archivePrefix  = "arXiv",
      primaryClass   = "hep-th",
      reportNumber   = "DAMTP-03-91",
      SLACcitation   = "%%CITATION = HEP-TH/0309180;%%"
}

@article{Poland:2018epd,
      author         = "Poland, David and Rychkov, Slava and Vichi, Alessandro",
      title          = "{The Conformal Bootstrap: Theory, Numerical Techniques,
                        and Applications}",
      year           = "2018",
      eprint         = "1805.04405",
      archivePrefix  = "arXiv",
      primaryClass   = "hep-th",
      SLACcitation   = "%%CITATION = ARXIV:1805.04405;%%"
}

@article{Weinberg:2012cd,
      author         = "Weinberg, Steven",
      title          = "{Minimal fields of canonical dimensionality are free}",
      journal        = "Phys. Rev.",
      volume         = "D86",
      year           = "2012",
      pages          = "105015",
      doi            = "10.1103/PhysRevD.86.105015",
      eprint         = "1210.3864",
      archivePrefix  = "arXiv",
      primaryClass   = "hep-th",
      reportNumber   = "UTTG-16-12",
      SLACcitation   = "%%CITATION = ARXIV:1210.3864;%%"
}

@article{Karateev:2019pvw,
      author         = "Karateev, Denis and Kravchuk, Petr and Serone, Marco and
                        Vichi, Alessandro",
      title          = "{Fermion Conformal Bootstrap in 4d}",
      journal        = "JHEP",
      volume         = "06",
      year           = "2019",
      pages          = "088",
      doi            = "10.1007/JHEP06(2019)088",
      eprint         = "1902.05969",
      archivePrefix  = "arXiv",
      primaryClass   = "hep-th",
      SLACcitation   = "%%CITATION = ARXIV:1902.05969;%%"
}

@article{Landry:2019qug,
	author = "Landry, Walter and Simmons-Duffin, David",
	title = "{Scaling the semidefinite program solver SDPB}",
	eprint = "1909.09745",
	archivePrefix = "arXiv",
	primaryClass = "hep-th",
	reportNumber = "CALT-TH 2019-038",
	month = "9",
	year = "2019"
}

@article{Simmons-Duffin:2012juh,
	author = "Simmons-Duffin, David",
	title = "{Projectors, Shadows, and Conformal Blocks}",
	eprint = "1204.3894",
	archivePrefix = "arXiv",
	primaryClass = "hep-th",
	doi = "10.1007/JHEP04(2014)146",
	journal = "JHEP",
	volume = "04",
	pages = "146",
	year = "2014"
}

@article{CastedoEcheverri:2015mkz,
	author = "Castedo Echeverri, Alejandro and Elkhidir, Emtinan and Karateev, Denis and Serone, Marco",
	title = "{Deconstructing Conformal Blocks in 4D CFT}",
	eprint = "1505.03750",
	archivePrefix = "arXiv",
	primaryClass = "hep-th",
	reportNumber = "SISSA-21-2015-FISI",
	doi = "10.1007/JHEP08(2015)101",
	journal = "JHEP",
	volume = "08",
	pages = "101",
	year = "2015"
}

@article{CastedoEcheverri:2016dfa,
	author = "Castedo Echeverri, Alejandro and Elkhidir, Emtinan and Karateev, Denis and Serone, Marco",
	title = "{Seed Conformal Blocks in 4D CFT}",
	eprint = "1601.05325",
	archivePrefix = "arXiv",
	primaryClass = "hep-th",
	reportNumber = "SISSA-02-2016-FISI",
	doi = "10.1007/JHEP02(2016)183",
	journal = "JHEP",
	volume = "02",
	pages = "183",
	year = "2016"
}

@article{Karateev:2026xxx,
	author         = "Karateev, Denis and Kravchuk, Petr ",
	title          = "{Automated Construction of Conformal Blocks in 4d CFTs}",
	journal = "in progress"
}

@article{Chang:2024whx,
	author = "Chang, Cyuan-Han and Dommes, Vasiliy and Erramilli, Rajeev S. and Homrich, Alexandre and Kravchuk, Petr and Liu, Aike and Mitchell, Matthew S. and Poland, David and Simmons-Duffin, David",
	title = "{Bootstrapping the 3d Ising stress tensor}",
	eprint = "2411.15300",
	archivePrefix = "arXiv",
	primaryClass = "hep-th",
	reportNumber = "CALT-TH 2024-047",
	doi = "10.1007/JHEP03(2025)136",
	journal = "JHEP",
	volume = "03",
	pages = "136",
	year = "2025"
}

@article{Chang:2025mwt,
    author = "Chang, Cyuan-Han and Dommes, Vasiliy and Kravchuk, Petr and Poland, David and Simmons-Duffin, David",
    title = "{Accurate bootstrap bounds from optimal interpolation}",
    eprint = "2509.14307",
    archivePrefix = "arXiv",
    primaryClass = "hep-th",
    reportNumber = "CALT-TH 2025-031",
    month = "9",
    year = "2025"
}

@article{Chester:2019ifh,
	author = "Chester, Shai M. and Landry, Walter and Liu, Junyu and Poland, David and Simmons-Duffin, David and Su, Ning and Vichi, Alessandro",
	title = "{Carving out OPE space and precise $O(2)$ model critical exponents}",
	eprint = "1912.03324",
	archivePrefix = "arXiv",
	primaryClass = "hep-th",
	reportNumber = "CALT-TH-2019-051",
	doi = "10.1007/JHEP06(2020)142",
	journal = "JHEP",
	volume = "06",
	pages = "142",
	year = "2020"
}

@article{Erramilli:2022kgp,
	author = "Erramilli, Rajeev S. and Iliesiu, Luca V. and Kravchuk, Petr and Liu, Aike and Poland, David and Simmons-Duffin, David",
	title = "{The Gross-Neveu-Yukawa archipelago}",
	eprint = "2210.02492",
	archivePrefix = "arXiv",
	primaryClass = "hep-th",
	reportNumber = "CALT-TH 2022-027",
	doi = "10.1007/JHEP02(2023)036",
	journal = "JHEP",
	volume = "02",
	pages = "036",
	year = "2023"
}

@article{Mitchell:2024hix,
	author = "Mitchell, Matthew S. and Poland, David",
	title = "{Bounding irrelevant operators in the 3d Gross-Neveu-Yukawa CFTs}",
	eprint = "2406.12974",
	archivePrefix = "arXiv",
	primaryClass = "hep-th",
	doi = "10.1007/JHEP09(2024)134",
	journal = "JHEP",
	volume = "09",
	pages = "134",
	year = "2024"
}

@article{Poland:2025ide,
	author = "Poland, David and Prilepina, Valentina and Tadi{\'c}, Petar",
	title = "{Mixed five-point correlators in the 3d Ising model}",
	eprint = "2507.01223",
	archivePrefix = "arXiv",
	primaryClass = "hep-th",
	reportNumber = "OUTP-21-11P",
	month = "7",
	year = "2025"
}

@article{El-Showk:2012cjh,
	author = "El-Showk, Sheer and Paulos, Miguel F. and Poland, David and Rychkov, Slava and Simmons-Duffin, David and Vichi, Alessandro",
	title = "{Solving the 3D Ising Model with the Conformal Bootstrap}",
	eprint = "1203.6064",
	archivePrefix = "arXiv",
	primaryClass = "hep-th",
	reportNumber = "LPTENS-12-07",
	doi = "10.1103/PhysRevD.86.025022",
	journal = "Phys. Rev. D",
	volume = "86",
	pages = "025022",
	year = "2012"
}

@article{Liu:2020tpf,
	author = "Liu, Junyu and Meltzer, David and Poland, David and Simmons-Duffin, David",
	title = "{The Lorentzian inversion formula and the spectrum of the 3d O(2) CFT}",
	eprint = "2007.07914",
	archivePrefix = "arXiv",
	primaryClass = "hep-th",
	doi = "10.1007/JHEP09(2020)115",
	journal = "JHEP",
	volume = "09",
	pages = "115",
	year = "2020",
	note = "[Erratum: JHEP 01, 206 (2021)]"
}

@article{Chester:2020iyt,
	author = "Chester, Shai M. and Landry, Walter and Liu, Junyu and Poland, David and Simmons-Duffin, David and Su, Ning and Vichi, Alessandro",
	title = "{Bootstrapping Heisenberg magnets and their cubic instability}",
	eprint = "2011.14647",
	archivePrefix = "arXiv",
	primaryClass = "hep-th",
	reportNumber = "CALT-TH-2020-053",
	doi = "10.1103/PhysRevD.104.105013",
	journal = "Phys. Rev. D",
	volume = "104",
	number = "10",
	pages = "105013",
	year = "2021"
}

@article{Reehorst:2024vyq,
	author = "Reehorst, Marten and Rychkov, Slava and Sirois, Benoit and van Rees, Balt C.",
	title = "{Bootstrapping frustrated magnets: the fate of the chiral ${\rm O}(N)\times {\rm O}(2)$ universality class}",
	eprint = "2405.19411",
	archivePrefix = "arXiv",
	primaryClass = "hep-th",
	doi = "10.21468/SciPostPhys.18.2.060",
	journal = "SciPost Phys.",
	volume = "18",
	number = "2",
	pages = "060",
	year = "2025"
}

@article{Reehorst:2021hmp,
	author = "Reehorst, Marten",
	title = "{Rigorous bounds on irrelevant operators in the 3d Ising model CFT}",
	eprint = "2111.12093",
	archivePrefix = "arXiv",
	primaryClass = "hep-th",
	doi = "10.1007/JHEP09(2022)177",
	journal = "JHEP",
	volume = "09",
	pages = "177",
	year = "2022"
}

@article{Reehorst:2020phk,
	author = "Reehorst, Marten and Refinetti, Maria and Vichi, Alessandro",
	title = "{Bootstrapping traceless symmetric $O(N)$ scalars}",
	eprint = "2012.08533",
	archivePrefix = "arXiv",
	primaryClass = "hep-th",
	doi = "10.21468/SciPostPhys.14.4.068",
	journal = "SciPost Phys.",
	volume = "14",
	number = "4",
	pages = "068",
	year = "2023"
}

@article{Dymarsky:2017xzb,
	author = "Dymarsky, Anatoly and Penedones, Joao and Trevisani, Emilio and Vichi, Alessandro",
	title = "{Charting the space of 3D CFTs with a continuous global symmetry}",
	eprint = "1705.04278",
	archivePrefix = "arXiv",
	primaryClass = "hep-th",
	doi = "10.1007/JHEP05(2019)098",
	journal = "JHEP",
	volume = "05",
	pages = "098",
	year = "2019"
}

@article{Manenti:2021elk,
	author = "Manenti, Andrea and Vichi, Alessandro",
	title = "{Exploring $SU(N)$ adjoint correlators in $3d$}",
	eprint = "2101.07318",
	archivePrefix = "arXiv",
	primaryClass = "hep-th",
	reportNumber = "UUITP-04/21",
	month = "1",
	year = "2021"
}

@article{He:2023ewx,
	author = "He, Yin-Chen and Rong, Junchen and Su, Ning and Vichi, Alessandro",
	title = "{Non-Abelian currents bootstrap}",
	eprint = "2302.11585",
	archivePrefix = "arXiv",
	primaryClass = "hep-th",
	doi = "10.1007/JHEP03(2024)175",
	journal = "JHEP",
	volume = "03",
	pages = "175",
	year = "2024"
}

@article{Reehorst:2019pzi,
	author = "Reehorst, Marten and Trevisani, Emilio and Vichi, Alessandro",
	title = "{Mixed Scalar-Current bootstrap in three dimensions}",
	eprint = "1911.05747",
	archivePrefix = "arXiv",
	primaryClass = "hep-th",
	doi = "10.1007/JHEP12(2020)156",
	journal = "JHEP",
	volume = "12",
	pages = "156",
	year = "2020"
}

@article{Bartlett-Tisdall:2024mbx,
	author = "Bartlett-Tisdall, Samuel and Herzog, Christopher P. and Schaub, Vladimir",
	title = "{An Atlas for 3d conformal field theories with a U(1) global symmetry}",
	eprint = "2412.01608",
	archivePrefix = "arXiv",
	primaryClass = "hep-th",
	doi = "10.1007/JHEP06(2025)237",
	journal = "JHEP",
	volume = "06",
	pages = "237",
	year = "2025"
}

@article{Cardy:1988cwa,
	author = "Cardy, John L.",
	title = "{Is There a c Theorem in Four-Dimensions?}",
	doi = "10.1016/0370-2693(88)90054-8",
	journal = "Phys. Lett. B",
	volume = "215",
	pages = "749--752",
	year = "1988"
}

@article{Osborn:1989td,
	author = "Osborn, H.",
	title = "{Derivation of a Four-dimensional $c$ Theorem}",
	reportNumber = "DAMTP-89-3",
	doi = "10.1016/0370-2693(89)90729-6",
	journal = "Phys. Lett. B",
	volume = "222",
	pages = "97--102",
	year = "1989"
}

@article{Jack:1990eb,
	author = "Jack, I. and Osborn, H.",
	title = "{Analogs for the $c$ Theorem for Four-dimensional Renormalizable Field Theories}",
	reportNumber = "DAMTP-90-02",
	doi = "10.1016/0550-3213(90)90584-Z",
	journal = "Nucl. Phys. B",
	volume = "343",
	pages = "647--688",
	year = "1990"
}

@article{Komargodski:2011vj,
	author = "Komargodski, Zohar and Schwimmer, Adam",
	title = "{On Renormalization Group Flows in Four Dimensions}",
	eprint = "1107.3987",
	archivePrefix = "arXiv",
	primaryClass = "hep-th",
	doi = "10.1007/JHEP12(2011)099",
	journal = "JHEP",
	volume = "12",
	pages = "099",
	year = "2011"
}

@article{Luty:2012ww,
	author = "Luty, Markus A. and Polchinski, Joseph and Rattazzi, Riccardo",
	title = "{The $a$-theorem and the Asymptotics of 4D Quantum Field Theory}",
	eprint = "1204.5221",
	archivePrefix = "arXiv",
	primaryClass = "hep-th",
	doi = "10.1007/JHEP01(2013)152",
	journal = "JHEP",
	volume = "01",
	pages = "152",
	year = "2013"
}

@misc{hyperion,
	author       = {Simmons-Duffin, David},
	title        = {Hyperion (Version 0.1.0.10)},
	howpublished = {\url{https://github.com/davidsd/hyperion}},
	note         = {Commit ec7c04ce935b6928cc3b1169b1035350446f710b}
}

@misc{blocks4d_haskell_gitlab,
	author       = {Kravchuk, Petr},
	title        = {blocks-4d-haskell (Version 0.1.0.0)},
	howpublished = {\url{https://gitlab.com/pkravchuk/blocks-4d-haskell}},
	note         = {Commit 4b5490e4aac0a47cde2bcbf29b23419b32aa77e8}
}

@misc{blocks4d_gitlab,
	author       = {Kravchuk, Petr},
	title        = {blocks-4d},
	howpublished = {\url{https://gitlab.com/pkravchuk/blocks-4d}},
	note         = {Commit bb00dcfd23670d80c6262d5a72e5a8b7e3868708}
}

@article{Chester:2025uxb,
	author = "Chester, Shai M. and Piazza, Alessandro and Reehorst, Marten and Su, Ning",
	title = "{Bootstrapping the Simplest Deconfined Quantum Critical Point}",
	eprint = "2507.06283",
	archivePrefix = "arXiv",
	primaryClass = "hep-th",
	month = "7",
	year = "2025"
}

@article{Li:2018lyb,
	author = "Li, Zhijin",
	title = "{Bootstrapping conformal QED3 and deconfined quantum critical point}",
	eprint = "1812.09281",
	archivePrefix = "arXiv",
	primaryClass = "hep-th",
	doi = "10.1007/JHEP11(2022)005",
	journal = "JHEP",
	volume = "11",
	pages = "005",
	year = "2022"
}

@article{Li:2021emd,
	author = "Li, Zhijin",
	title = "{Conformality and self-duality of Nf=2 QED3}",
	eprint = "2107.09020",
	archivePrefix = "arXiv",
	primaryClass = "hep-th",
	doi = "10.1016/j.physletb.2022.137192",
	journal = "Phys. Lett. B",
	volume = "831",
	pages = "137192",
	year = "2022"
}

@article{Chester:2023njo,
	author = "Chester, Shai M. and Su, Ning",
	title = "{Bootstrapping Deconfined Quantum Tricriticality}",
	eprint = "2310.08343",
	archivePrefix = "arXiv",
	primaryClass = "hep-th",
	doi = "10.1103/PhysRevLett.132.111601",
	journal = "Phys. Rev. Lett.",
	volume = "132",
	number = "11",
	pages = "111601",
	year = "2024"
}

@article{He:2021xvg,
	author = "He, Yin-Chen and Rong, Junchen and Su, Ning",
	title = "{A roadmap for bootstrapping critical gauge theories: decoupling operators of conformal field theories in $d>2$ dimensions}",
	eprint = "2101.07262",
	archivePrefix = "arXiv",
	primaryClass = "hep-th",
	reportNumber = "DESY-21-123",
	doi = "10.21468/SciPostPhys.11.6.111",
	journal = "SciPost Phys.",
	volume = "11",
	pages = "111",
	year = "2021"
}

@article{He:2021sto,
	author = "He, Yin-Chen and Rong, Junchen and Su, Ning",
	title = "{Conformal bootstrap bounds for the $U(1)$ Dirac spin liquid and $N=7$ Stiefel liquid}",
	eprint = "2107.14637",
	archivePrefix = "arXiv",
	primaryClass = "cond-mat.str-el",
	reportNumber = "DESY-21-124",
	doi = "10.21468/SciPostPhys.13.2.014",
	journal = "SciPost Phys.",
	volume = "13",
	number = "2",
	pages = "014",
	year = "2022"
}

@article{DiPietro:2020jne,
	author = "Di Pietro, Lorenzo and Serone, Marco",
	title = "{Looking through the QCD Conformal Window with Perturbation Theory}",
	eprint = "2003.01742",
	archivePrefix = "arXiv",
	primaryClass = "hep-th",
	doi = "10.1007/JHEP07(2020)049",
	journal = "JHEP",
	volume = "07",
	pages = "049",
	year = "2020"
}

@article{Ciccone:2024guw,
	author = "Ciccone, Riccardo and De Cesare, Fabiana and Di Pietro, Lorenzo and Serone, Marco",
	title = "{Exploring confinement in Anti-de Sitter space}",
	eprint = "2407.06268",
	archivePrefix = "arXiv",
	primaryClass = "hep-th",
	doi = "10.1007/JHEP12(2024)218",
	journal = "JHEP",
	volume = "12",
	pages = "218",
	year = "2024",
	note = "[Erratum: JHEP 06, 037 (2025)]"
}

@article{Ciccone:2025dqx,
	author = "Ciccone, Riccardo and De Cesare, Fabiana and Di Pietro, Lorenzo and Serone, Marco",
	title = "{QCD in AdS}",
	eprint = "2511.04752",
	archivePrefix = "arXiv",
	primaryClass = "hep-th",
	month = "11",
	year = "2025"
}

@article{DiPietro:2025ozw,
	author = "Di Pietro, Lorenzo and Kousvos, Stefanos R. and Meineri, Marco and Piazza, Alessandro and Serone, Marco and Vichi, Alessandro",
	title = "{A Bootstrap Study of Confinement in AdS}",
	eprint = "2512.00150",
	archivePrefix = "arXiv",
	primaryClass = "hep-th",
	month = "11",
	year = "2025"
}

@article{Beneke:1997qd,
	author = "Beneke, M. and Braun, Vladimir M. and Kivel, N.",
	title = "{Large order behavior due to ultraviolet renormalons in QCD}",
	eprint = "hep-ph/9703389",
	archivePrefix = "arXiv",
	reportNumber = "CERN-TH-97-050, CERN-TH-97-50, NORDITA-97-17-P",
	doi = "10.1016/S0370-2693(97)00562-5",
	journal = "Phys. Lett. B",
	volume = "404",
	pages = "315--320",
	year = "1997"
}

@article{Bauer:1997gs,
	author = "Bauer, Christian W. and Manohar, Aneesh V.",
	title = "{Renormalization group scaling of the 1/m**2 HQET Lagrangian}",
	eprint = "hep-ph/9708306",
	archivePrefix = "arXiv",
	reportNumber = "UCSD-PTH-97-19, UTPT-97-17",
	doi = "10.1103/PhysRevD.57.337",
	journal = "Phys. Rev. D",
	volume = "57",
	pages = "337--343",
	year = "1998"
}

@article{Baikov:2014qja,
	author = {Baikov, P. A. and Chetyrkin, K. G. and K{\"u}hn, J. H.},
	title = "{Quark Mass and Field Anomalous Dimensions to ${\cal O}(\alpha_s^5)$}",
	eprint = "1402.6611",
	archivePrefix = "arXiv",
	primaryClass = "hep-ph",
	reportNumber = "SFB-CPP-14-16, TTP14-007",
	doi = "10.1007/JHEP10(2014)076",
	journal = "JHEP",
	volume = "10",
	pages = "076",
	year = "2014"
}

@article{Herzog:2017ohr,
	author = "Herzog, F. and Ruijl, B. and Ueda, T. and Vermaseren, J. A. M. and Vogt, A.",
	title = "{The five-loop beta function of Yang-Mills theory with fermions}",
	eprint = "1701.01404",
	archivePrefix = "arXiv",
	primaryClass = "hep-ph",
	reportNumber = "NIKHEF-2017-001, LTH-1117",
	doi = "10.1007/JHEP02(2017)090",
	journal = "JHEP",
	volume = "02",
	pages = "090",
	year = "2017"
}

@article{Hasenfratz:2018wpq,
	author = "Hasenfratz, Anna and Rebbi, Claudio and Witzel, Oliver",
	title = {{Determination of the N$_f$=12 step scaling function using M{\"o}bius domain wall fermions}},
	eprint = "1810.05176",
	archivePrefix = "arXiv",
	primaryClass = "hep-lat",
	doi = "10.22323/1.334.0306",
	journal = "PoS",
	volume = "LATTICE2018",
	pages = "306",
	year = "2019"
}

@article{Fodor:2018uih,
	author = "Fodor, Zoltan and Holland, Kieran and Kuti, Julius and Nogradi, Daniel and Wong, Chik Him",
	title = "{Is SU(3) gauge theory with 13 massless flavors conformal?}",
	eprint = "1811.05024",
	archivePrefix = "arXiv",
	primaryClass = "hep-lat",
	doi = "10.22323/1.334.0198",
	journal = "PoS",
	volume = "LATTICE2018",
	pages = "198",
	year = "2018"
}

@article{Hasenfratz:2019dpr,
	author = "Hasenfratz, Anna and Rebbi, Claudio and Witzel, Oliver",
	title = "{Gradient flow step-scaling function for SU(3) with twelve flavors}",
	eprint = "1909.05842",
	archivePrefix = "arXiv",
	primaryClass = "hep-lat",
	doi = "10.1103/PhysRevD.100.114508",
	journal = "Phys. Rev. D",
	volume = "100",
	number = "11",
	pages = "114508",
	year = "2019"
}

@article{Hasenfratz:2024fad,
	author = "Hasenfratz, Anna and Peterson, Curtis T.",
	title = "{Infrared fixed point in the massless twelve-flavor SU(3) gauge-fermion system}",
	eprint = "2402.18038",
	archivePrefix = "arXiv",
	primaryClass = "hep-lat",
	reportNumber = "FERMILAB-PUB-24-0098-V",
	doi = "10.1103/PhysRevD.109.114507",
	journal = "Phys. Rev. D",
	volume = "109",
	number = "11",
	pages = "114507",
	year = "2024"
}

@phdthesis{Manenti:2020nyv,
	author = "Manenti, Andrea",
	title = "{Bootstrapping superconformal field theories in four dimensions}",
	doi = "10.5075/epfl-thesis-8330",
	school = "Ecole Polytechnique, Lausanne",
	year = "2020"
}

@article{Lin:2019vgi,
	author = "Lin, Ying-Hsuan and Meltzer, David and Shao, Shu-Heng and Stergiou, Andreas",
	title = "{Bounds on Triangle Anomalies in (3+1)d}",
	eprint = "1909.11676",
	archivePrefix = "arXiv",
	primaryClass = "hep-th",
	reportNumber = "CALT-TH-2019-034, LA-UR-19-30323",
	doi = "10.1103/PhysRevD.101.125007",
	journal = "Phys. Rev. D",
	volume = "101",
	number = "12",
	pages = "125007",
	year = "2020"
}

@misc{CREATE,
  author       = {{King's College London}},
  title        = {King's Computational Research, Engineering and Technology Environment ({CREATE})},
  year         = {2024},
  howpublished = {\url{https://doi.org/10.18742/rnvf-m076}},
  note         = {Retrieved Nov 13, 2024}
}

\end{document}